# AUTOMATED NANOSIMS MEASUREMENTS OF SPINEL STARDUST FROM THE MURRAY METEORITE

FRANK GYNGARD,[1,*] ERNST ZINNER,[1] LARRY R. NITTLER,[2] ALAIN MORGAND,[3]

FRANK J. STADERMANN,[1] and K. MAIRIN HYNES[1]

[1]Laboratory for Space Sciences and the Department of Physics, Washington University, One

Brookings Drive, St. Louis, MO 63130, USA, fgyngard@dtm.ciw.edu

[2]Department of Terrestrial Magnetism, Carnegie Institution of Washington, 5241 Broad Branch

Road, NW Washington, DC 20015, USA

[3]Cameca, 29 Quai des Grésillons, Gennevilliers, 92622 Cedex, France

[*]Present address: Department of Terrestrial Magnetism, Carnegie Institution of Washington,

5241 Broad Branch Road, NW Washington, DC 20015, USA

Short Title:

AUTOMATED MEASUREMENTS OF SPINEL STARDUST

# ABSTRACT

We report new O isotopic data on 41 presolar oxide grains, 38 $MgAl_2O_4$ (spinel) and 3 $Al_2O_3$ from the CM2 meteorite Murray, identified with a recently developed automated measurement system for the NanoSIMS. We have also obtained Mg-Al isotopic results on 29 of the same grains (26 spinel and 3 $Al_2O_3$). The majority of the grains have O isotopic compositions typical of most presolar oxides, fall well into the four previously defined groups, and are most likely condensates from either red giant branch or asymptotic giant branch stars. We have also discovered several grains with more unusual O and Mg compositions suggesting formation in extreme astrophysical environments, such as novae and supernovae. One of these grains has massive enrichments in $^{17}O$, $^{25}Mg$, and $^{26}Mg$, which are isotopic signatures indicative of condensation from nova ejecta. Two grains of supernova origin were also discovered: one has a large $^{18}O/^{16}O$ ratio typical of Group 4 presolar oxides; another grain is substantially enriched in $^{16}O$, and also contains radiogenic $^{44}Ca$ from the decay of $^{44}Ti$, a likely condensate from material originating in the O-rich inner zones of a Type II supernova. In addition, several Group 2 presolar spinel grains also have large $^{25}Mg$ and $^{26}Mg$ isotopic anomalies that are difficult to explain by standard nucleosynthesis in low-mass stars. Auger elemental spectral analyses were performed on the grains and qualitatively suggest that presolar spinel may not have higher-than-stoichiometric Al/Mg ratios, in contrast to SIMS results obtained here and reported previously.

*Subject headings:* dust, extinction – novae, cataclysmic variables – nuclear reactions, nucleosynthesis, abundances – stars: AGB and post AGB –supernovae: general

# 1. INTRODUCTION

Presolar grains are astrophysical fossils contained in primitive meteorites and provide valuable insight into stellar evolution and the formation of the elements. These grains are literally stardust which condensed out of the windy envelopes of ancient stars and ejecta from supernova (SN) explosions; they have provided some of the starting material for the formation of the Solar System (SS). Stardust grains represent a nucleosynthetic snapshot of their parent stars at the time of grain condensation and afford insight into many astrophysical processes (Clayton & Nittler 2004; Meyer & Zinner 2006). First identified over two decades ago by their unusual isotopic compositions when compared to solar or terrestrial values, presolar grains have provided constraints on models of nucleosynthesis in asymptotic and red giant branch (AGB and RGB) stars and supernovae (SNe), dust grain condensation conditions in stellar environments, and the chemical evolution of the galaxy (GCE).

While carbonaceous presolar grains (such as nanodiamonds, SiC, and graphite) were discovered first (Amari et al. 1990; Bernatowicz et al. 1987; Lewis et al. 1987), O-rich phases, including alumina ($Al_2O_3$), spinel ($MgAl_2O_4$), hibonite ($CaAl_{12}O_{19}$), and silicates, have also since been identified. Corundum is the stable mineral phase of $Al_2O_3$ on Earth; however, as microstructural analyses indicate that presolar $Al_2O_3$ can exist in other forms (Stroud et al. 2004), we prefer the terminology "alumina" or "$Al_2O_3$." For a comprehensive account of the various O-rich presolar grain types and their discovery, see Zinner (2007). Because of its refractory nature, an ability to better retain trace elements in its crystal lattice, and typically larger (often several μm) grain sizes compared to presolar oxides (usually ≲ 1 μm), SiC is the best studied presolar grain type, and roughly 10,000 SiC grains have been individually analyzed for their isotopic compositions. In contrast, the study of presolar oxides is only just beginning to approach a similar degree of maturity, for several reasons. Because the SS is O-rich (C/O ~ 1/2 – Lodders 2003) and meteorites possess high abundances of oxide mineral phases, finding *presolar* oxides in extraterrestrial samples among literally thousands of isotopically normal grains (mainly those of SS origin) surrounding them, requires time consuming and often inefficient techniques. This situation is unlike the case for SiC, in which chemical separation treatments can produce residues enriched (> 90%) in presolar SiC (Amari, Lewis, & Anders 1994). Traditionally, two techniques have been successfully applied to identify O-rich presolar grains in meteoritic separates. Oxygen isotopic raster imaging has proven efficient for presolar grain discovery in samples containing dense and tightly packed grains (see Fig. 1 of Nguyen, Zinner, & Lewis 2003), prepared as either polished meteoritic thin sections or size-separated aggregates. In this way, thousands of particles can be measured. Oxygen-rich presolar grains can subsequently be seen as anomalous O isotopic "hotspots" in the ion images when compared to surrounding grains. Alternatively, when oxide rich grain residues have been produced by chemical and physical processing that provide well-separated grains on the sample mounts after deposition, individual (manual) analyses can be performed with little or no isotopic dilution from adjacent particles, thereby better obtaining the true composition of the grains. With the implementation of the NanoSIMS (a high-spatial-resolution, multi-detection secondary ion mass spectrometer), the analysis of tightly packed grain separates has the advantage that if grain density is high enough, presolar grains can be readily discovered (Nguyen & Zinner 2004); however, it has its drawbacks. Due to primary beam overlap onto nearby solar-composition, or "normal", grains, the true isotopic compositions of the anomalous grains may be diluted and the magnitude of the isotopic anomalies become lower limits. This isotopic dilution is more pronounced for grains with depletions in $^{17}O$ and/or $^{18}O$ which are undercounted (Nguyen et al. 2007) and, because the grains are typically embedded in the matrix of surrounding grains, further measurements of other isotopic systems can be compromised (or may be impossible). This dilution can also affect grains enriched in $^{17}O$ and $^{18}O$ (although not as much) – see Figure 5 of Zinner and Gyngard (2009). Measurements of individual grains have been facilitated by the development of techniques (Gröner & Hoppe 2006; Nittler & Alexander 2003) in which well-dispersed grains are automatically identified from ion images by computer algorithms that recognize and define, based on customizable parameters such as ion intensity or grain size, discrete particles in the analyzed areas. Electrostatic deflection of the primary ion beam (under computer control) onto the identified grains then allows their isotopic compositions to be obtained in a systematic, automated manner. Combined with the high sensitivity at high-mass resolution and multi-detection capabilities of the NanoSIMS, this method of automatic analysis has been shown to be particularly useful for searches of low-abundant presolar grain types (Heck et al. 2006).

Based on their O isotopic composition, most presolar oxide grains have been loosely assigned to four groups (Nittler et al. 1997), thought to reflect origins in distinct types of stars. Although the boundaries between the groups are not well defined, the groups have been roughly established as: Group 1 grains, which are characterized by enrichments, often large, in $^{17}O$ and moderate depletions in $^{18}O$; Group 2 grains, which have slight enrichments in $^{17}O$ (comparable to those of most Group 1 grains) but more extreme $^{18}O$ depletions than Group 1 grains; Group 3 grains, which are enriched in $^{16}O$; and Group 4 grains, which are enriched in either $^{18}O$ or both of the heavy O isotopes. The isotopic signatures of most Group 1-3 grains are consistent with model calculations of the envelope compositions of low-mass ($M \lesssim 3M_\odot$), close-to solar metallicity RGB and AGB stars (e.g. Boothroyd, Sackmann, & Wasserburg 1994), while the enigmatic Group 4 grains have been theorized to possibly be SN condensates (Choi et al. 1998;

Nittler et al. 2008).  Prior to the studies by Zinner et al. (2005) and Nittler et al. (2008), few measurements of isotopic systems other than O had been primarily reported for presolar $Al_2O_3$ (Nittler et al. 1997) and hibonite (Choi, Wasserburg, & Huss 1999).  Unlike $Al_2O_3$ grains, spinel and hibonite grains provide the opportunity to determine isotopic compositions of major elements besides O; in fact, roughly half of the spinel grains measured for Mg isotopes exhibit substantial isotopic anomalies (both enrichments and/or depletions in $^{25}$Mg and $^{26}$Mg relative to $^{24}$Mg) generally consistent with origins inferred from their O isotopic compositions (Zinner et al. 2005).  However, some previously discovered presolar spinel grains have extreme Mg compositions (enrichments in both $^{25}$Mg and $^{26}$Mg over 1000‰) that are difficult to reconcile with their measured O-isotopic compositions in the context of current AGB nucleosynthesis models (Iliadis et al. 2008; Nittler et al. 2008).  It also should be made clear here that excesses in $^{25}$Mg and $^{26}$Mg are effectively the same as depletions in $^{24}$Mg, and vice versa.

In this work, we report the development of an automated, high-mass-resolution system for the analysis of individual grains in the NanoSIMS and its first application to a spinel-rich residue of grains from the Murray CM2 carbonaceous chondrite.  We have also obtained the Mg-Al isotopic compositions of a subset of the grains, expanding the database of presolar spinel grains analyzed for Mg isotopes roughly twofold, and for several grains, Ca and Ti isotopes were also measured.  In addition, we obtained Auger elemental spectra on the majority of the grains, in order to compare them to our results from SIMS analysis.  Reports of parts of this work have been presented previously (Gyngard et al. 2009; Gyngard et al. 2010; Gyngard & Zinner 2009; Zinner & Gyngard 2009).

## 2. EXPERIMENTAL TECHNIQUES

The pedigree of the Murray residue analyzed here is described in detail by Tang and Anders (1988). Essentially, a 12.7 g fragment of the meteorite was chemically treated with HF and HCl in order to remove both silicates and carbonaceous material and the resulting residue was size separated into three fractions. The CG fraction, used in this study, has an average grain diameter of 0.45 μm and predominantly consists of spinel. The procedure for mounting these grains for NanoSIMS analysis is described by Zinner et al. (2003).

In collaboration with Cameca, the manufacturer of the NanoSIMS, we have developed an automated, high-mass-resolution measurement technique for the rapid analysis of large numbers of well-separated grains, similar to previously developed systems on NanoSIMS and ims-6f microprobes (Gröner & Hoppe 2006; Nittler & Alexander 2003). For this study, automated measurements consisted of the following sequence: a 20 x 20 μm$^2$ area was presputtered with a high-current primary $Cs^+$ beam to remove surface contamination and to implant Cs in the grains present, which enhances the negative secondary ion yield. Scanning ion images of $^{16}O^-$, $^{17}O^-$, $^{18}O^-$, $^{24}MgO^-$, and $AlO^-$ were subsequently acquired in multi-detection with a ~1 pA, 100 nm $Cs^+$ primary beam. With particle definition software (Nittler 1996), individual grains were identified in the $^{16}O^-$ images, sorted according to various criteria (such as proximity to surrounding grains), and individually measured by electrostatically deflecting the primary beam onto each grain and rastering over a square area with a side approximately twice the grain's diameter (as determined from the ion images). After analysis of each selected grain in a given imaged area, the sample stage was moved and the entire process repeated. The viability and reproducibility of the system will be discussed further below; however, it is clear that with this technique, thousands of grains can be automatically analyzed over the course of several days, with little user intervention. The automated system we have developed here differs from a previous system for the NanoSIMS (Gröner & Hoppe 2006) in that it is completely integrated into the instrument's software, allowing for greater measurement flexibility and compatibility. Image processing is accomplished with software written in the IDL language (ITT); information is passed between the main NanoSIMS software and the IDL program via text files. It must be noted that one grain, C4-8, was not found by the automatic grain technique, but rather was observed as an $^{17}O^-$ "hotspot" in a large pile of grains in an automatically obtained 20 x 20 $\mu m^2$ image. Because of its close proximity to other material, this grain was not selected by the grain recognition algorithm to be automatically measured in that area; its O isotopic composition was measured manually.

For the present study, the practical measurement conditions (arrived at by trying to minimize needless grain erosion, while optimizing for efficient identification of stardust grains) were as follows: each area scanned was subjected to 2 minutes of pre-sputtering to clean off surface contamination and implant Cs in the surface of the grains to enhance secondary ion yields, followed by a 5.5 minute acquisition of the ion image used for particle identification. On average, each area contained 10 identified grains whose individual measurements took roughly 1.5 minutes each. These conditions led to a net throughput of ~ 2.25 minutes per particle, not including instrumental setup and tuning, and resulted in about 640 individual particle measurements per day. At present, the system automatically determines the integration time of each measurement based on the approximate size of the grain to be measured; however, in the future, efficiency could be increased if measurement times were adjusted for desired statistical precision. In addition, increasing the number of grains per image by optimizing the density of the deposited grains on the mounts would also increase the overall efficiency. Unfortunately, for liquid suspensions of oxides, as opposed to, for example, SiC, the grains have a tendency to clump together, making the production of well-dispersed (with minimum distances between individual grains of several microns) sample mounts difficult.

Synthetic $Al_2O_3$ grains, with essentially normal O isotopic compositions, were used for normalization of the O data; however, this is not absolutely necessary as presolar grains are identified by their large isotopic deviations from bulk averages of meteoritic grains. The $Al_2O_3$ standard in this study is the same as that used by Zinner et al. (2003) and was shown to have approximately SMOW (standard mean ocean water: $^{17}O/^{16}O = 0.0003829$ and $^{18}O/^{16}O = 0.0020052$) isotopic composition. Once grains were identified as isotopically anomalous and hence presolar, they were analyzed with the PHI 700 Scanning Auger Nanoprobe at Washington University Nanoprobe, equipped with a field-emission electron source, in order to relocate the grains, take high-resolution SEM images, and independently determine the grains' mineralogy via Auger electron spectroscopy. While, in principle, standard energy dispersive X-ray (EDX) analysis could obtain the same information, the fact that many of the grains were clumped together made Auger analysis preferable. Due to the short inelastic mean free path of Auger electrons (only electrons emitted from the very top surface escape the sample with their original Auger energy), analysis of the energy distribution of emitted Auger electrons results in a smaller *effective* analysis depth of Auger electrons emitted with characteristic energies (a few nm) than that for X-rays (~1 μm) (Stadermann et al. 2009). This makes it possible to determine elemental compositions with less contribution from adjacent or underlying material, in particular for grains that are ≲ 200 nm in size and are touched or encompassed by adjacent grains. The spinel grains were analyzed with a 10 kV, 10 nA electron beam. Unlike silicates (Floss & Stadermann 2009), in which substantial electron beam damage can alter grain chemistry, the spinel grains showed no evidence of beam damage under these conditions. In most cases, spectra were acquired by rastering the electron beam over part of the grain and collecting Auger electrons with energies from 30 to 1730 eV. In a few cases, elemental maps of selected elements (e.g., O, Mg, Al, Si) were also taken of areas in which the grain density was particularly high and the full spectral measurements yielded inconclusive spectra, mostly due to sample stage drift when the analysis area moved during the short analysis of some small grains. These elemental maps provide a qualitative overview of the elemental distribution and can be very helpful in the identification of interfering adjacent grains.

In roughly a quarter of the grains, it was apparent from the SEM images (Figure 1a, b) that the "particle" defined by the grain recognition software and measured automatically in the NanoSIMS consisted, in fact, of multiple grains. In most cases, a "manual operator" would likely have observed this and excluded the grains from measurement. In order to obtain the true isotopic composition of the actual presolar grain, we obtained raster-ion images of the three O isotopes with a ~100 nm $Cs^+$ primary beam in the NanoSIMS (Figure 1c, d). After identification of the anomalous grain, we carefully sputtered away the surrounding material using the NanoSIMS' $Cs^+$ primary beam (Zinner & Gyngard 2009). This process allowed us to not only determine the O composition of such "cleaned up" grains without interferences, but also enhanced our ability to make future measurements of the isotopic compositions of other elements without contributions from adjacent normal grains. This is especially important as subsequent isotopic analyses of these samples with positive secondary ions (e.g., for Mg, K, Ca, Ti) require an $O^-$ primary beam, which has a much larger beam spot (500 – 1000 nm versus 100 nm) than the $Cs^+$ beam, thereby making the measurements more susceptible to contamination due to beam overlap. If enough material of the presolar grain remained, we re-measured the O isotopes after removal of nearby material; however, in a few cases, we extracted the O isotopic ratios from the ion images themselves, to preserve as much of the grain as possible for further analyses.

For grains that had not been completely sputtered away during O isotopic analysis, we also measured their Mg isotopic composition and Al content with the NanoSIMS. This was done by simultaneous detection of secondary ions of the three stable Mg isotopes ($^{24}Mg^+$, $^{25}Mg^+$, and $^{26}Mg^+$) and $^{27}Al^+$, produced by bombardment with a ~10 pA, ~ 0.5 – 1 μm O$^-$ primary beam rastered over the grain. Solar system spinel grains – as determined from their approximately normal (see discussion below) O isotopic composition – nearby on the sample mount were used as standards for normalizing the Mg isotopes and determining the Al$^+$/Mg$^+$ relative sensitivity factor (RSF), needed to quantify Al/Mg elemental ratios. Two particularly interesting grains, discussed below, were further characterized in a separate measurement session of $^{39}K^+$, $^{41}K^+$, $^{40}Ca^+$, $^{42}Ca^+$, $^{43}Ca^+$, $^{44}Ca^+$, $^{47}Ti^+$, and $^{48}Ti^+$ with a combination of multidetection and magnetic peak switching. Beam blanking was employed in order to eliminate contamination due to beam overlap onto nearby material. A perovskite ($CaTiO_3$) standard was used for normalization of the Ca-Ti isotopic compositions and a RSF of 2.83 between Ca and Ti (determined from a NBS 610 glass) was assumed.

### 3. RESULTS

#### 3.1 O Isotopes

A total of 312 20 x 20 μm$^2$ areas were automatically mapped and 3152 grains were identified and measured for their O isotopic composition. The average isotopic composition of the 357 non-anomalous grains with errors less than 20‰, measured in multiple automated runs over the course of a week, is δ$^{17}$O/$^{16}$O = -47 ± 26‰ and δ$^{18}$O/$^{16}$O = -59 ± 19‰ (errors are the standard deviations of the 357 grain measurements), consistent with previously reported spinel grains from Ca-Al rich inclusions (CAIs) having compositions of $\delta^{17}O = \delta^{18}O = -40‰$ (Clayton 1993), even though we cannot absolutely state that all of these grains are, in fact, spinel and not $Al_2O_3$. These results indicate an about 2-3% uncertainty in the automated technique – good enough for efficient identification of presolar grains. The closeness of the secondary ion extraction optics to the sample surface (400 μm) in the NanoSIMS likely increases the sensitivity of the automatic system to sample height topography effects compared to other ion microprobes. For grains not remeasured for their O isotopes, the errors on the O isotopic ratios are calculated by including uncertainties from both counting statistics and the overall scatter on standards measured during the automated run in which the grains were discovered (not the 3% overall uncertainty quoted above).

Forty-one grains (Table 1) were found to have anomalous O isotopic compositions, defined by enrichments or depletions more than 3σ away from the average O composition of nearby "solar" meteoritic spinel grains, thus corresponding to a presolar number abundance of ~ 1.3% among the grains of the Murray CG residue. The O isotopic ratios, shown in Figure 2, are similar to those obtained by previous analyses on presolar oxide grains and fall into all four previously identified groups (Nittler 1997). As typical for presolar oxides, the majority of the grains are enriched in $^{17}O$, characteristic of Group 1 and 2 grains. Six grains exhibit large $^{18}O$ depletions ($^{18}O/^{16}O < 10^{-3}$) characteristic of Group 2 grains, in addition to at least one Group 3 grain (depletions in both $^{17}O$ and $^{18}O$) and one Group 4 grain (enriched in $^{18}O$). Grain C4-8 is unique in that its $^{17}O/^{16}O$ ratio of 4.4 x $10^{-2}$ is the largest ever observed in presolar O-rich grains and is at least an order of magnitude greater than the theoretical maximum value in models of low-mass AGB stars, thought to be the progenitors of Group 1 grains. One other grain, T54 (Nittler et al. 1997), has been reported to have a comparably large enrichment in $^{17}O$ ($^{17}O/^{16}O =$ 1.4 x 10$^{-2}$), though still a factor of 3 less than that of C4-8; such large enrichments most likely indicate a nova origin for these grains, the astrophysical implications of which will be discussed below.

As mentioned above, SEM images showed that some of the grains measured automatically actually consisted of two or more grains tightly packed together. In such cases, oxygen isotopic ion imaging (for example, Figures 1c, d) proved very effective at establishing which of the grains were indeed presolar. The O isotopic composition of the grain depicted in Figure 1 (12-13-3) is shown in Figure 2, as well as the compositions of grains measured both automatically and manually (following careful sputter removal of the normal grains nearby). For the majority of the grains, the shift of their isotopic composition after sputter removal is subtle. For at least two grains, however, the compositions were shown to exhibit much more extreme anomalies than those obtained by the automated analysis. Grain 7-5-7, which was originally thought to be a fairly typical Group 3 grain, is, in fact, extremely enriched in $^{16}$O ($^{17}$O/$^{16}$O = 3.9 x 10$^{-5}$ and $^{18}$O/$^{16}$O = 2.1 x 10$^{-4}$) and likely not of an AGB star origin. The Group 4 grain from this study, grain 12-13-3, was shown to be substantially more enriched in both $^{17}$O and $^{18}$O than the automatic measurement revealed (Gyngard et al. 2009).

### 3.2 Mg, Ca, and Ti Isotopes

The Mg isotopic compositions of the grains are included in Table 1 and shown in Figure 3. In general, the Mg results are similar to those of previous analyses of presolar spinel grains (Zinner et al. 2005). Approximately half of the grains cluster around normal Mg isotopic composition; however, the rest exhibit considerable isotopic anomalies. The most extreme Mg anomalous grains are 14-12-7 (a Group 2 grain) and C4-8, with $\delta^{25}$Mg = 1034 ± 8‰ and 949 ± 8‰, $\delta^{26}$Mg = 1040 ± 8‰ and 929 ± 8‰, respectively (delta notation: $\delta^i$X/$^j$X = [($^i$X/$^j$X)$_{measured}$/(

$^i$X/$^j$X)$_\odot$ - 1] x 1000); these are some of the largest $^{25}$Mg excesses observed to date. There is a cluster of six grains exhibiting similar Mg isotopic ratios, characterized by large $^{26}$Mg excesses (200 – 400‰) and $^{25}$Mg depletions of about 200‰, but dissimilar O isotopic compositions. The most $^{16}$O rich grain, 7-5-7, is unique in having essentially normal $^{26}$Mg/$^{24}$Mg, but a $^{25}$Mg depletion of about 300‰.

After the Mg-Al measurements, two grains had sufficient material remaining for analysis of their K, Ca, and Ti isotopic compositions. For grain 7-5-7, no counts of $^{42}$Ca and $^{43}$Ca were recorded, and its $^{41}$K/$^{39}$K and $^{47}$Ti/$^{48}$Ti ratios are normal within (admittedly large) errors. Due to the small size of this grain (very little material remained after the Mg-Al measurement) and the fact that K, Ca, and Ti are present only as trace elements, analytical uncertainties are large. However, the grain has a clear enrichment in $^{44}$Ca ($^{44}$Ca/$^{40}$Ca = 1.3 ± 0.4) corresponding to a $^{44}$Ca excess of ~58,000‰. This leads to an inferred $^{44}$Ti/$^{48}$Ti ratio of (3.6 ± 0.8) x 10$^{-3}$. Although we tried to avoid nearby contamination, any terrestrial Ti would reduce the calculated $^{44}$Ti/$^{48}$Ti ratio, and, in fact, such dilution of the grain's true ratio may have occurred, as we observed an unexpectedly high $^{48}$Ti$^+$/$^{40}$Ca$^+$ ratio, perhaps from material left over from a nearby grain. However, an explanation for why there would be significantly more Ti than Ca contamination is unknown. Unfortunately, the second grain, C4-8, contained too little Ca and Ti for obtaining a meaningful isotopic analysis – all isotopic ratios are normal within very large uncertainties.

### 3.3 Al/Mg Elemental Ratios

Based on our NanoSIMS measurements of Al and Mg, the presolar spinel grains of this study appear to have elevated Al/Mg ratios (shown in Table 1, where measured) compared to the stoichiometric atomic ratio of two. This has been observed before in studies of presolar Murray spinel (Zinner et al. 2005) and has been hypothesized to possibly be indicative of formation in a high temperature environment (Simon et al. 1994). While it is tempting to try to quantify the Auger elemental spectra taken by using the Al/Mg ratios determined for the nearby solar spinel grains as standards, the fact that some spinel-like minerals from CAIs from the Murchison CM2 chondrite exhibit elevated Al/Mg ratios, indicative of high temperature alteration (Simon et al. 1994), precludes the assumption of stoichiometric composition. In addition, most of the spectra are extremely noisy due to the small size of the grains, which in turn makes any absolute quantification of the RSF between Al and Mg inherently more uncertain. Qualitatively, however, there seems to be little or no difference in the Al and Mg compositions of the Auger spectra of solar and non-solar spinel. The question of whether or not most presolar Murray spinel grains are truly non-stoichiometric or whether these results are artifacts of the instrumental techniques will be discussed below in Section 4.4.

## 4. DISCUSSION

The astrophysical setting, nucleosynthesis, and mixing processes in RGB and AGB stars, along with comparisons to presolar grain data, have been extensively reviewed (Busso, Gallino, & Wasserburg 1999; Nittler et al. 2008); however, a few points relevant to the discussion here will be presented. In Figure 4, a schematic diagram is shown of the various astrophysical processes that are primarily responsible for the O isotopic compositions of both Group 1 and 2 grains. During main-sequence core H-burning, $^{17}O$ is enriched due to proton capture on $^{16}O$ in the CNO cycle, greatly increasing the $^{17}O/^{16}O$ ratio, while $^{18}O$ is essentially destroyed due to the $^{18}O(p,\alpha)^{15}N$ reaction operating at relatively low (~1.5 x $10^7$ K) temperatures (Boothroyd et al. 1994). After the cessation of H-burning in their core, low-to-intermediate mass (< 8$M_\odot$) stars cool and expand, moving up the RGB in the Hertzprung-Russell diagram. Immediately subsequent to this expansion, these stars undergo a deep convection event (1st dredge-up), in which the remaining products of H-burning in layers including those that experience CNO cycling are mixed up to the stellar surface, largely homogenizing the envelope of the star and enriching the surface in previously synthesized $^{17}$O-rich and $^{18}$O-poor material. The surface $^{18}$O/$^{16}$O ratio is reduced by ~20% and the $^{17}$O/$^{16}$O ratio increases to a degree that depends primarily on the initial mass of the star (which effectively determines the maximum depth of the 1st dredge-up). The predicted $^{17}$O/$^{16}$O ratio increases with stellar mass up to a maximum of 0.004 for a 2.5$M_\odot$ star, after which the $^{17}$O/$^{16}$O ratio begins to decline with increasing mass (Boothroyd & Sackmann 1999). It should be noted here that stars of mass greater than 3.5$M_\odot$ undergo another deep mixing event, the 2nd dredge-up, which does not significantly alter the O isotopic composition of the star's surface (Lattanzio & Boothroyd 1997). Other mixing processes reaching sites experiencing nuclear reactions, occurring later in the evolution of the star during the RGB and AGB phase, can also occur, e.g., hot bottom burning (HBB) for intermediate-mass stars and cool bottom processing (CBP) for low-mass stars, but they primarily destroy $^{18}$O and have only a small effect on the star's $^{17}$O/$^{16}$O ratio; these processes will be discussed in more detail later.

Within this general context, most grain data fall well into the previously established groups, and indicate an origin in low-mass AGB or RGB stars of lower-than- or close-to-solar metallicity. Consistent with previous observations, the Group 2 grains must have experienced CBP to account for their low $^{18}$O/$^{16}$O ratios. However, some of the grains found in this study do not fit within this framework, and require different astrophysical origins. One grain (C4-8), with the largest $^{17}$O/$^{16}$O ratios discovered to date, probably condensed from nova ejecta and is further discussed in Section 4.1. Several Group 2 grains have extreme Mg isotopic compositions which are difficult to produce in models of single low-mass AGB stars, perhaps indicative of a binary star origin. These grains are discussed in Section 4.2. The one Group 4 grain we have identified has an isotopic composition consistent with formation in SN ejecta, similar to previous results for Group 4 grains. Grain 7-5-7, with its large $^{16}$O enrichment and clear evidence for $^{44}$Ti, also likely condensed from SN material. These two grains' isotopic compositions are compared with SN model predictions in Section 4.3. In Section 4.4, we make some general remarks addressing the question of the grains' stoichiometry, and discuss whether or not Al/Mg ratios as determined by SIMS measurements for presolar spinel are conclusive.

### 4.1 Nova Nucleosynthesis and Grain Production

#### 4.1.1 Classical Novae

Classical novae are stellar explosions in close binary systems caused by the buildup of H-rich material onto the white dwarf (WD) companion of a cool, red, main sequence (observationally often a K or M type) star. The angular momentum of accrued material from the RGB companion star causes the formation of an accretion disk around the WD. This H-rich matter eventually falls onto the WD, and as the material at the base of the star's envelope is very dense (1000 – 10,000 g cm$^{-3}$), it is essentially electron degenerate. Degenerate material cannot maintain hydrostatic equilibrium (e.g., pressure will not increase with temperature) because it does not expand and cool when the temperature increases. Through continuing mass-transfer episodes from the companion star, temperatures increase and nuclear reactions, mainly with H, can be powered at the base of the WD's envelope without any subsequent expansion, leading to a thermonuclear runaway (TNR) on the surface of the WD. Degeneracy ceases when the Fermi temperature is reached and the envelope can then expand, thus ending the explosion and returning the WD to its pre-outburst state; however, the continuing accretion of H-rich material from the companion RGB star typically leads to further episodes of TNR and the corresponding ejection of nuclear processed matter (José, Hernanz, & Iliadis 2006). Novae ejecta are usually dominated by C/O ratios much less than one (except for very massive explosions) and typically occur for WDs that are carbon-oxygen (CO) or oxygen-neon (ONe) rich, depending on the main-sequence mass of the star that evolved to become the WD. The secondary star in the binary system can itself re-accrete material ejected from these nova outbursts, possibly affecting its surface isotopic composition (Marks, Sarna, & Prialnik 1997), although the gross compositional change may be small. Nuclear processes affecting the O and Mg isotopes are dominated by proton capture reactions. The significant production of $^{17}$O in CO novae is most sensitive to the $^{17}$O(p,$\alpha$) reaction; whereas for ONe novae, variations in the $^{18}$F(p,$\alpha$) and $^{17}$F(p,$\alpha$) reactions can greatly affect the $^{16}$O, $^{17}$O, and $^{18}$F (and thereby $^{18}$O) abundances. Uncertainties in many reaction rates can strongly affect predictions for the production of the Mg isotopes in both CO and ONe nova models. In particular, the rate for $^{23}$Na(p,$\gamma$) strongly affects $^{24,25,26}$Mg, and $^{26}$Al$^g$(p,$\gamma$), which is responsible for production of $^{26}$Al – though, in general, CO nova models produce considerably less $^{26}$Al than do ONe models (Iliadis et al. 2002).

### 4.1.2 Presolar Grains from Novae

Principally through proton captures, novae are efficient producers of the stable isotopes $^{13}$C, $^{15}$N, and $^{17}$O, as well as the radioactive isotopes $^{22}$Na and $^{26}$Al (José & Hernanz 1998; Kovetz & Prialnik 1997; Starrfield et al. 1998). To date, primarily carbonaceous phases of presolar grains (e.g., SiC and graphite), exhibiting a combination of low $^{12}$C/$^{13}$C and $^{14}$N/$^{15}$N ratios, high $^{30}$Si/$^{28}$Si, and high inferred $^{26}$Al/$^{27}$Al and low $^{20}$Ne/$^{22}$Ne ratios (when measured), have been purported to have formed in novae (Amari et al. 2001; Heck et al. 2007; José et al. 2004). There is some controversy, however, as a SN origin cannot absolutely be ruled out, based on a large $^{47}$Ti excess in one grain and large $^{28}$Si, $^{49}$Ti, and $^{44}$Ca isotopic excesses in another putative nova grain candidate (Nittler & Hoppe 2005). Also, quantitative comparison of the grain data and nova models indicates that the model compositions are much more extreme than those of the grains (Amari et al. 2001); hence, the grains must have condensed in nova ejecta that were mixed with a substantial amount of close-to-solar composition matter. Despite this uncertainty, one recently discovered SiC grain with a $^{12}$C/$^{13}$C ratio of about 1 (Nittler, Alexander, & Nguyen 2006) most likely formed in ONe nova ejecta. In addition, José and Hernanz (2007) recently concluded that $^{26}$Al/$^{27}$Al ratios cannot be used to distinguish between a SN and nova origin, and that most of the Amari et al. (2001) grains are likely of nova origin; though, clearly more multi-element isotopic studies of nova-candidate SiC grains are needed to unambiguously determine their origins. Prior to this study, only one possible nova oxide ($Al_2O_3$) grain, T54, has been identified (Nittler et al. 1997). Its extreme $^{17}$O enrichment ($^{17}$O/$^{16}$O = 1.41 x 10$^{-2}$) cannot be produced by standard AGB star nucleosynthesis and the grain has been suggested to be a probable CO nova condensate (Nittler & Hoppe 2005).

### 4.1.3 Grain C4-8

The O isotopic composition of grain C4-8 of this study is characterized by a huge enrichment in $^{17}$O, with $^{17}$O/$^{16}$O = (4.40 ± 0.01) x 10$^{-2}$. Although Group 1 grains also have $^{17}$O excesses (Zinner 2007), RGB and AGB stars, the most likely sources of these grains (Nittler 1997), cannot produce such a high $^{17}$O/$^{16}$O ratio. In other words, the large $^{17}$O excesses observed in grains C4-8 and T54 (i.e., $^{17}$O/$^{16}$O >> 4 x 10$^{-3}$) cannot be explained by typical nucleosynthesis processes that take place in the parent stars of Group 1 grains. In nova explosions, maximum temperatures greater than 10$^8$ K can be reached (depending on the WD mass), causing fast CNO processing (i.e., in the "hot" CNO cycle), or non-equilibrium burning, in which the rate of nuclear processing is predominately governed by $\beta^+$-decay half-lives (Pagel 2009; Starrfield, Iliadis, & Hix 2008; Starrfield et al. 1972). Quantitative predictions for the mean mass-averaged O isotopic composition of CO and ONe nova models for different WD masses (José et al. 2004) are shown in Figure 5, along with the compositions of grains C4-8 and T54. In some cases, multiple compositions have been calculated for a given model, such as for the $0.8M_\odot$ and $1.15M_\odot$ CO models, corresponding to different amounts of mixing between the solar-like accreted envelope and core WD material (José & Hernanz 1998). Unlike in the case of putative SiC nova grains, ONe novae cannot be the progenitors of grains C4-8 and T54, as ONe novae produce $^{18}O/^{16}O$ ratios larger than solar; the best match to the O isotopic compositions of these grains is produced by CO nova models. Model CO8 of José et al. (2004), a $1.15M_\odot$ CO model computed for an initial composition with a 50% degree of mixing between core material and accreted envelope, best explains the composition of grain C4-8; however, a $0.8 M_\odot$ model (CO4) cannot unequivocally be ruled out. Coincidentally, a $1.25 M_\odot$ CO model calculated by Yaron et al. (2005) predicts an $^{17}O/^{16}O$ of 0.049, also in reasonable agreement with the composition of C4-8; however, the $^{18}O/^{16}O$ ratio is not explicitly given in their work. The O isotopic composition of grain T54 is very well explained by a $0.8M_\odot$ CO model (CO2), also calculated and evolved from an initial composition of 50% core material and 50% envelope. It is interesting to note that models CO2 and CO8 have been computed with an updated (increased) $^{18}F$+p reaction rate (Hernanz et al. 1999), as opposed to the other CO models, leading to significantly less production of $^{18}O$ compared to models of the same WD mass, regardless of the amount of core-envelope mixing. The O isotopic compositions of both C4-8 and T54 seem to favor the models calculated with the updated rate.

Grain C4-8 is significantly enriched in both $^{25,26}$Mg ($\delta^{25}$Mg = 949 ± 9‰ and $\delta^{26}$Mg = 929 ± 7‰) and, similar to the situation for the O isotopes, nucleosynthesis in RGB and AGB stars cannot produce the observed Mg isotopic excesses. During the main sequence stage of close-to-solar metallicity low-mass ($\lesssim$ 3M$_\odot$) stars, temperatures in the zones containing material brought to the surface by 1st dredge-up are not high enough for Mg to undergo any substantial nuclear processing; therefore, predicted enrichments are only up to a few permil (Karakas & Lattanzio 2003). During the AGB phase, $^{25}$Mg and $^{26}$Mg can be created in the H-burning shell by successive p-captures when the Mg-Al chain (of which $^{26}$Al is the main product) is activated, as well as by the $^{22}$Ne($\alpha$,n)$^{25}$Mg and $^{22}$Ne($\alpha$,$\gamma$)$^{26}$Mg reactions operating in the He-burning shell. Third dredge-up increases the amount of heavy Mg isotopes in the envelope. However, models of AGB stars of close-to-solar metallicity do not predict $^{25}$Mg enrichments greater than ~ 40‰ (Zinner et al. 2005). Larger enhancements are possible for lower metallicity stars (up to ~ 200‰); however, these stars are assumed to have initial $^{25}$Mg/$^{24}$Mg ratios lower than solar, severely limiting the maximum $\delta^{25}$Mg values that can be reached. The production of the isotopes of Mg and Al will be discussed in more detail later in the context of Group 2 grains, but, regardless, $^{25}$Mg/$^{24}$Mg ratios greater than 300‰ cannot be produced in any single-AGB-star scenario.

Extreme Mg isotopic anomalies and production of $^{26}$Al expected in nova outbursts and the Mg composition of C4-8 are shown, along with the nova model predictions of José et al. (2004) for Mg and Al, in Figure 6. For grains like C4-8 with large anomalies in both $^{25}$Mg and $^{26}$Mg, it is difficult to disentangle $^{26}$Mg directly produced by nucleosynthesis and radiogenic $^{26}$Mg from the decay of $^{26}$Al. Thus for the nova models plotted in Figure 6, contributions of the decay of $^{26}$Al to $^{26}$Mg have been included. Because the Al/Mg ratio in stoichiometric spinel is 2, whereas the solar ratio is 0.082 (Lodders 2003), the [26]Al model abundances have been multiplied by a factor of 24, close to the value used by Lugaro et al. (2007).

In contrast to the case of the O isotopes, the Mg isotopic composition of C4-8 is very close to that predicted for the 0.6 $M_\odot$ CO nova model (CO1). Except for the 0.6 $M_\odot$ model, both CO and ONe novae are predicted to produce enormous enrichments of both [25]Mg and [26]Mg (whether or not radiogenic [26]Mg is included) relative to solar abundances. In model CO1, only little nuclear processing (i.e., destruction) of [24]Mg occurs, as the WD mass is too low to significantly activate the [24]Mg(p,γ)[25]Al reaction (José et al. 2004). Unfortunately, model CO8, which gives the best fit to the O composition of C4-8, predicts Mg isotopic ratios much too extreme to explain the grain data. Ultimately, all nova models (CO or ONe), except for the unique case of CO1, produce [25]Mg/[24]Mg and [26]Mg/[24]Mg ratios too high by at least an order of magnitude to fit the grain's composition. On the other hand, model CO1 predicts [17]O/[16]O and [18]O/[16]O ratios significantly lower than that of C4-8. No Mg isotopic ratios are reported for the Yaron et al. (2005) model, only elemental mass fractions for Z > 8, thus no comparisons with the grain data can be made. As it was completely consumed during its O isotopic measurement, the Mg and Al compositions of grain T54 were not determined, and thus no additional information is available to confirm the grain's tentative nova origin, or to constrain model predictions.

The presence of intermediate-mass elements from Si to Ca has been astronomically observed in the dust shells of some classical novae (Andreä, Drechsel, & Starrfield 1994), and model calculations of ONe novae predict nucleosynthetic production of Si through Ca (José, Coc, & Hernanz 2001). For very violent nova explosions with large progenitor masses, recent model calculations have indicated nucleosynthesis beyond Ca, which is typically the endpoint of nova nucleosynthesis, and predictions for the Ti compositions in some ONe nova models have shown extreme enrichments in $^{46}$Ti, $^{47}$Ti, and $^{49}$Ti – up to roughly 1000% for $^{46}$Ti (José & Hernanz 2007). Unfortunately, measurement of the isotopic compositions of Ca and Ti of C4-8 cannot help to further constrain the origin of this unique grain, as statistically significant isotopic results could not be obtained. Unlike the situation for the supposed nova SiC grains, both C4-8 and T54 are likely to come from CO novae, which are not expected to undergo nuclear burning of elements (such as Si, S, Ca, Ti, etc.) heavier than those effected by CNO and Mg-Al cycle processing. Due to the dearth of seed elements heavier than O for which nuclear burning (typically proton capture) can occur, the scope of nucleosynthesis that can happen for elements heavier than those involved in CNO and Mg-Al cycling is limited. While spinel grains afford the opportunity to accurately measure their O, Mg, and Al compositions (all major elements), the Auger spectrum of C4-8 indicated little or no Ca (for which Auger analysis is particularly sensitive) or Ti, making isotopic analysis difficult because of poor counting statistics and susceptibility to possible contamination on the sample mount.

### 4.1.4 Condensation Conditions of Novae Grains

The mineralogy of condensed grains typically depends on the C/O ratio, as most of the C and O tends to get tied up in the very stable CO molecule (Lodders & Fegley 1995). Excess of either O or C typically allows the formation of oxide/silicate or carbonaceous phases, respectively. Despite this condition, equilibrium condensation of SiC and graphite can occur in ONe novae due to the presence of heavier elements (e.g., Mg, Al, Si, etc.) that alter the condensation chemistry (José & Hernanz 2007; José et al. 2004), and C-rich dust has actually been observed around both CO and ONe novae (Gehrz et al. 1998). Although we cannot assume grain survival probabilities are the same for oxides and carbonaceous grains, in light of the fact that nova ejecta are O-rich, it remains puzzling why relatively few oxide nova grains have been identified compared to the number of purported carbonaceous nova grains.

It is outside the scope of this paper to discuss in detail the genesis of all candidate nova grains (both C-rich and O-rich) and compare them with nucleosynthesis predictions. However, by comparing the number of presolar nova grains discovered so far among their grain types, we can gain insight into the relative abundances of C- or O-rich nova grains preserved in primitive meteorites. In addition to oxide grains C4-8 and T54 discussed earlier, one presolar silicate grain with a possible nova origin has also been identified (Nguyen et al. 2007). This grain has an $^{17}O/^{16}O$ ratio of $(9.54 \pm 0.11) \times 10^{-3}$, which is significantly larger than the maximum $^{17}O/^{16}O$ ratio of $4 \times 10^{-3}$ that can be achieved in a single $2.5M_\odot$ RGB star. A search of the Presolar Database (Hynes & Gyngard 2009) revealed an additional four oxide grains that have $^{17}O/^{16}O$ ratios greater than $4 \times 10^{-3}$ by more than $2\sigma$, suggesting a possible nova origin for these grains as well (Choi et al. 1999; Nguyen et al. 2003; Nittler & Hoppe 2005) – these grains were also discussed in Nittler et al. (2008) as being likely nova condensates. Based on the total number of presolar oxide and silicate grains in the Presolar Database (as of writing, 612 and 230 grains, respectively) and by assuming between two and six oxide grains to have a nova origin, we estimate that nova grains constitute ~0.3 – 1% of presolar oxide grains and 0.4% of presolar silicate grains, corresponding to 0.4 – 0.8% of all O-rich grains identified so far. As mentioned earlier, SiC is the most studied type of presolar grain, with almost 10,000 individual grains measured. Of these grains, at least eight have been identified as having a possible nova origin (Amari et al. 2001; Gao & Nittler 1997; Heck et al. 2007; Hoppe et al. 1996; Nittler & Alexander 2003; Nittler & Hoppe 2005), although in principal, most of these grains could have either a nova or SN origin (Nittler & Hoppe 2005). If all eight of these grains did originate in novae, they represent ~0.09% of the SiC population.  However, this is an upper limit; searches often target only specific rare grain types and other grain types, especially mainstream SiC (~93% of the SiC population), are either not measured or not reported in the literature.  This bias is most obvious for SiC X grains, which constitute only ~1% of presolar SiC grains, but account for ~5% of SiC grains analyzed.  To correct for this overabundance of rare grain types that have been measured, we extrapolated from the number of X grains in the Presolar Database and the reported SiC grain type abundances (Hoppe & Ott 1997; Meyer & Zinner 2006; Nittler & Alexander 2003) to calculate the likely fraction of presolar SiC grains from novae that would have resulted from unbiased surveys.  This yields a bias corrected value of only ~0.02% of SiC grains originating in novae and this percentage would be even lower if some of the eight SiC grains assumed to be from novae actually condensed in SN ejecta.  It is therefore likely that the smaller number of oxide nova grains than SiC nova grains that have been found is due to statistics: over 10 times more SiC grains have been measured than oxide grains and almost forty times more SiC grains than silicate grains.  After accounting for this measurement bias, experimental grain data tend to show that the relative fraction of nova grains among O-rich grains is several times, up to an order of magnitude, higher than the fraction of SiC grains from novae.  These considerations are at least qualitatively consistent with astronomical observations of nova dust shells exhibiting enhancements in O-rich minerals.  Although the small number of nova grains of any grain type that have been measured, as well as the fundamental uncertainty in the specificity of their origin, puts large uncertainties on these abundance estimates, it is clear that SiC nova grains do not largely outnumber O-rich nova grains.  In addition, the maximal abundances of presolar oxides and silicates are at least 100 ppm and 180 ppm, respectively, in primitive meteorites – greater than the 55 ppm maximum observed for SiC in minimally altered

CR meteorites (Davidson et al. 2009), with the caveat that these abundances are not determined for the same meteorites. Since primitive meteorites are the closest representatives of the SS as a whole, the total abundance of O-rich nova grains in the SS, and thus probably interstellar space, must be much higher than that of SiC grains from novae. Further dedicated searches for O-rich grains showing the signature of nova explosions are needed to confirm these results, and, the discovery of more of these grains will provide crucial information about nucleosynthesis in binary star systems.

### 4.2 Group 2 grains

As mentioned above, Group 2 presolar oxide and silicate grains are defined by large depletions in $^{18}$O and modest enrichments in $^{17}$O. In addition, they typically have high inferred $^{26}$Al/$^{27}$Al ratios. While the star's initial mass essentially determines the maximum surface $^{17}$O/$^{16}$O ratio that can be reached by the first dredge-up, the corresponding $^{18}$O/$^{16}$O ratio is reduced by only a relatively minor amount (up to 20%). In principle, at least two different possibilities (or a combination thereof) can explain the low $^{18}$O/$^{16}$O ratios seen in the Group 2 grains: (1) the grains originated in low metallicity stars, hence with lower-than-solar initial $^{18}$O/$^{16}$O ratios (Boothroyd & Sackmann 1999), or (2) some extra mixing process, such as HBB in intermediate-mass AGB stars or CBP in low-mass AGB stars, that cycled material through H-burning regions, resulted in the destruction of $^{18}$O, reducing the $^{18}$O/$^{16}$O ratio. The low $^{18}$O/$^{16}$O ratios of the Group 2 grains (and the inferred metallicities of their parent stars) almost solely reflect the effect of mixing processes; thus, it is impossible to directly infer metallicity from the O isotopic ratios. If grains that are on the border between Groups 1 and 2 are excluded, it can be seen in Figure 2 that the observed width of the distribution of the $^{17}$O/$^{16}$O ratios for Group 2 grains is at least a factor of 5 smaller than that for Group 1 grains; hence these grains must have condensed from stars with a much narrower range of initial stellar masses than the Group 1 grains – this is discussed in detail in Nittler et al. (2008).

### 4.2.1 AGB Stars and Extra Mixing Processes

After the cessation of core H- and He-burning, low- and intermediate-mass stars enter the AGB phase of their lives. At this point, the star consists of a CO electron-degenerate core, surrounded by a convective envelope. Between the convective envelope and core are two thin layers, separated by a He intershell of mass $10^{-2} – 10^{-3}$ $M_\odot$ (Straniero et al. 1997), that alternately burn H and He. For most of the time during the AGB phase, the star burns H; however, as the He abundance and electron degeneracy grow in the intershell, He begins to burn at its bottom and a thermonuclear runaway occurs (the thermal pulse, or TP), making the intershell convective and shutting down H burning. Depending on the mass of the star, He-burning can last for as long as 100 years for low-mass stars ($\lesssim$ 4 $M_\odot$) to only 10 years for more massive stars. After the TP, the convective envelope dips down into part of the intershell (third dredge-up, or TDU) and this process mixes the freshly synthesized products of He burning and previous H burning, as well as neutron-capture nucleosynthesis, into the envelope, and H burning begins anew. It should be remarked here, however, that the TP-AGB phase of the star does not significantly alter the O isotopic composition of the envelope.

During the TP-AGB phase, stars of mass roughly 4 – 7 $M_\odot$ undergo HBB, in which the bottom of the star's convective envelope extends into the top of the H-burning shell where CNO cycling occurs. Large depletions in $^{18}O$ are predicted by HBB models, yielding $^{18}O/^{16}O$ ratios as low as $10^{-7}$ (Boothroyd, Sackmann, & Wasserburg 1995). Such extreme ratios are not observed in the grains, though it is possible that nearby terrestrial material on the sample mount could dilute the measured O and shift $^{18}O/^{16}O$ ratios, particularly for grains with extreme $^{18}O$ depletions, toward solar-like values. HBB also efficiently converts C into N, so that the star maintains C/O < 1 in the envelope (whereas, typically, TDU increases the amount of C such that the star fairly rapidly becomes a C-star), allowing for the continued condensation of O-rich grains. One extremely $^{18}$O depleted spinel grain, OC2 (Zinner et al. 2005), with extreme $^{25}$Mg and $^{26}$Mg enrichments of 433‰ and 1170‰, respectively, was suggested to have formed in an intermediate-mass AGB star (Lugaro et al. 2007; Zinner et al. 2005), with the assumption of a small amount of dilution of its original composition during measurement. However, re-evaluation of the rate of the $^{16}$O(p,$\gamma$)$^{17}$F reaction (Iliadis et al. 2008), responsible for production of $^{17}$O, limits the minimum $^{17}$O/$^{16}$O ratio obtained in the envelope of stars experiencing HBB to 2.5 x 10$^{-3}$, effectively ruling out an intermediate-mass star as the progenitor of this grain, which has an $^{17}$O/$^{16}$O ratio of 1.25 ± 0.07 x 10$^{-3}$. Furthermore, since, as discussed above, the $^{17}$O/$^{16}$O ratio is largely an indicator of initial stellar mass, the $^{17}$O/$^{16}$O ratios of most of the Group 2 grains indicate an origin in low-mass (1.2 – 1 .6 M$_\odot$) AGB stars, which do not experience HBB. However, as the $^{17}$O/$^{16}$O ratio decreases with stellar mass above 2.5M$_\odot$, stars of 6 – 8M$_\odot$ are also predicted to produce a spread of $^{17}$O/$^{16}$O ratios consistent with that of the Group 2 grains. Regardless, an examination of the currently available grain data (Hynes & Gyngard 2009) so far shows no evidence of any grains from intermediate-mass stars, consistent with the conclusions reached by Nittler (2009).

As has been pointed out previously by many authors, the best explanation for the O and Al isotopic compositions of the Group 2 grains is CBP (or "extra mixing") that might occur in low-mass AGB stars (Boothroyd & Sackmann 1999; Nollett, Busso, & Wasserburg 2003; Wasserburg, Boothroyd, & Sackmann 1995). This process is characterized by an as yet not well understood physical process in which material is slowly circulated from the star's convective envelope down to near the H-burning shell, where it undergoes some moderate nuclear processing before being cycled back up into the envelope.  Despite uncertainty in the exact mechanism of circulation (see Busso et al. (2007) for an updated discussion on the possible physical mechanisms), low $^{12}C/^{13}C$ and $^{18}O/^{16}O$ ratios in astronomical observations of RGB stars (e.g., Kahane et al. 1992) and laboratory isotopic measurements of presolar grains (Nittler et al. 2008; Zinner et al. 2006) show substantial evidence that CBP occurs in both RGB and AGB stars.  Also, at least for RGB stars, three-dimensional hydrodynamics simulations have self-consistently producing deep mixing from Rayleigh-Taylor instability above the H-burning shell (Eggleton 2006).  In their parameterized model, Nollet et al. (2003) adjusted the maximum temperature the circulated material reaches (affecting the $^{26}Al/^{27}Al$ ratio) and the circulation rate (primarily affecting the $^{18}O/^{16}O$ ratio) and were able to explain the O and Al isotopic compositions of most Group 2 grains.  In fact, most of the additional O data of the Group 2 grains from this study are generally consistent with previous results, indicating that the grains' parent stars must have experienced some degree of CBP.

### 4.2.2 Extreme Mg Isotopic Compositions of Some Group 2 Grains

The Mg isotopic compositions – in particular $^{25}Mg$ excesses greater than ~150 ‰ – of at least two of the Group 2 grains (13-30-9 and 14-12-7) are too extreme to be explained by models of low-mass AGB stars with C/O < 1.  Larger enhancements in $^{25}Mg$ are predicted when the star becomes C-rich; however, it is probably not possible to form spinel in such an environment.  At least three other Group 2 spinel grains have been discovered with similar Mg isotopic anomalies, M16 (Zinner & Gyngard 2009), OC2, and UOC-S2, as well as 3 presolar hibonite grains (Nittler et al. 2008).  The very large $^{25}Mg$ and $^{26}Mg$ isotopic enrichments in these grains are well fitted by models of Mg production, particularly when we include contributions from radiogenic $^{26}Mg$ from the decay of $^{26}$Al, in intermediate-mass (2.25 – 6 M$_\odot$) AGB stars (Karakas & Lattanzio 2003); however, as we have already pointed out, this source has been effectively ruled out by the O isotope data. Recently, Nittler et al. (2008) suggested that mass transfer in a binary star system, from an intermediate-mass AGB star enriched in $^{25}$Mg and $^{26}$Mg to a low-mass companion, could, in principle, explain the compositions of these grains. However, detailed calculations need to be performed in order to see if such a mass transfer scenario can simultaneously reproduce the $^{17}$O/$^{16}$O, $^{18}$O/$^{16}$O, $^{25}$Mg/$^{24}$Mg, and $^{26}$Mg/$^{24}$Mg ratios in these grains. At present, their origin remains uncertain.

### *4.3 Unique O-rich Grains from Supernovae*

As mentioned before, most Group 4 grains have been theorized to be SN condensates. Choi et al. (1998) were first to suggest that the isotopic composition of a Group 4 grain could be reproduced by mixing appropriate amounts of material from various layers of massive stars exploding as SNe. While a detailed analysis of nucleosynthesis in SNe is outside the scope of this paper, we follow the typical prescription of dividing the SN ejecta in zones (or layers) based on the most abundant elements in that zone – see Meyer et al. (1995) for a primer. For the discussion here, it is most relevant that the O-rich inner zones of SNe are dominated by $^{16}$O, whereas the He-burning shell (He/C zone) is heavily enriched in $^{18}$O (with C/O > 1). Mixing of the outer H envelope material, which has undergone only moderate nuclear processing, with a small amount of the He/C layer in the correct proportions can produce the observed $^{18}$O/$^{16}$O ratios of the Group 4 grains. In addition, at least one extreme Al$_2$O$_3$ grain, T84 (Nittler et al. 1998), with a massive enrichment in $^{16}$O, and a unique silicate grain, B10A (Messenger, Keller, & Lauretta 2005), exhibiting a large depletion in $^{17}$O and huge enrichment in $^{18}$O, are also undoubtedly of a SN origin.

Although we have discovered a second extremely $^{16}$O enhanced grain that likely derives its O isotopic composition from the O-rich inner zones of a SN, it still remains puzzling why so few grains of similar composition have been identified. Approximately 10% of oxide grains have $^{18}$O enrichments and their isotopic compositions are generally consistent with the suggestion of Choi et al. (1998) that these grains are SN condensates forming from a mixture of $^{18}$O material from the He/C zone with material from the outer H envelope. Perhaps some as yet unknown destruction mechanism or chemical constraint inhibits the successful condensation of oxide grains with significant amounts of material from the O-rich inner zones. Quite possibly, dust condensing from inner zone material may be predominately small (< 50nm) in size and easily destroyed by sputtering in reverse shocks (Nozawa et al. 2007) or, alternatively, less efficiently detected by currently available experimental techniques.

Two anomalous spinel grains from this study likely condensed in SN ejecta: grain 7-5-7 has an O isotopic composition very similar to that of T84 (Figure 7) and grain 12-13-3 is a fairly extreme Group 4 grain, with $^{17}$O and $^{18}$O excesses larger than those typically observed in most Group 4 grains. Grain 7-5-7 has a very low $^{25}$Mg/$^{24}$Mg ratio ($\delta^{25}$Mg = -276 ± 10‰) and an approximately solar $^{26}$Mg/$^{24}$Mg ratio ($\delta^{26}$Mg = 13 ± 12‰). Calculations performed by a simple mixture of different amounts of material from the He/C zone with an average of the O-rich inner zones for a 15 M$_\odot$ SN model (Rauscher et al. 2002) can very well reproduce the O isotopic compositions of grains 7-5-7 and T84 (Figure 7); however, the same mixture fails to explain the Mg and and Ca-Ti isotopic compositions of grain 7-5-7 (T84 was only analyzed for O isotopes). Both the O/Ne and O/C zones are typically characterized by large $^{25}$Mg/$^{24}$Mg and $^{26}$Mg/$^{24}$Mg ratios, in contrast to the composition of grain 7-5-7. In the O/C zone, $^{25}$Mg and $^{26}$Mg enrichments are far too extreme (with enhancements of up to ~40,000‰ relative to solar) to have the grain acquire any contribution from this zone (see Figure 8). Rather, it likely must have incorporated material from a mixture of the O/Si and O/Ne zones. In general, the Mg isotopic effects completely dominate contributions from $^{26}$Al in the O-rich zones (though substantial amounts are found there), and only the He/N zone has a high $^{26}$Al/$^{27}$Al ratio. While astronomical observations have shown evidence of heterogeneity in SN ejecta (Hughes et al. 2000), quantitative multi-zone mixing calculations (Yoshida & Hashimoto 2004) often fail to reproduce the isotopic compositions of multiple elemental systems in the same SN grain. Therefore, it is not surprising, that granular mixing of more SN layers is required to fit the multiple element isotope data of grain 7-5-7. Mixing of variable amounts of SN ejecta from the O/Si zone, O/Ne zone, He/C zone, He/N zone, and the H envelope of the 15M$_\odot$ model by Rauscher et al. (2002) is able to reproduce the isotopic ratios of grain 7-5-7 to about 5 - 15%, but overproduce the $^{44}$Ti/$^{48}$Ti ratio by at least a factor of 2.7 (Figure 9). New model simulations of a 15M$_\odot$ SN by Woosley and Heger (2007) yield much better agreement between the predictions and grain data, and, although the fit is slightly worse for the O isotopic composition, the measured and calculated $^{44}$Ti/$^{48}$Ti ratios agree within the one sigma analytical error and the inherent uncertainties of the model simulations themselves. This mixture only fits the Mg isotopic composition by including the contribution of radiogenic $^{26}$Mg from the decay of $^{26}$Al, yielding an inferred $^{26}$Al/$^{27}$Al ratio of 2.8 x $10^{-2}$, consistent with values observed previously for presolar grains from SNe (e.g., Group 4 grains and SiC X grains).

In contrast to the Mg isotopic composition of grain 7-5-7, grain 12-13-3 has enrichments in both $^{25}$Mg and $^{26}$Mg, with $\delta^{25}$Mg = 71± 13‰ and $\delta^{26}$Mg = 735 ± 16‰ (Figure 3). The O isotopic compositions and $^{26}$Mg enrichment can be best explained by the mixing of He/C zone, He/N zone, and H envelope material; however, the $^{25}$Mg excess requires material from either the O/C zone or O/Ne zone (or both), which are rich in $^{16}$O. Mixtures calculated by including contributions from all these necessary zones can roughly reproduce grain 12-13-3's $^{17}$O/$^{16}$O and $^{18}$O/$^{16}$O ratios, as well as the $\delta^{25}$Mg and $\delta^{26}$Mg values (if radiogenic $^{26}$Mg from the decay of $^{26}$Al in the He/N zone is included, with $^{26}$Al/$^{27}$Al ratios on the order of $10^{-2}$ to $10^{-3}$ depending upon the exact mixture). It should be noted that the mixing calculations performed here do not attempt to find any unique or preferential mixing scheme to match the grain data (for example, by minimizing the $\chi^2$ distribution), but demonstrate that, at least in principle, such mixing can reproduce the grains' isotopic compositions.

### 4.4 Stoichiometry of Presolar Spinel Grains

Almost all SIMS studies of previously discovered presolar spinel grains analyzed for Mg and Al have reported non-stoichiometric Al/Mg ratios, both greater than and less than 2 (Choi et al. 1998; Nittler et al. 2008; Nittler et al. 1994; Zinner et al. 2005). Likewise, the Al/Mg ratios of the grains analyzed in this study (Table 1) also show higher-than-stoichiometric values, when normalized to nearby isotopically normal spinel grains on the sample mount. Elevated Al/Mg ratios in presolar spinel could provide important information about the stellar environment in which they condensed, as high Al/Mg ratios observed in spinels from CAIs have been attributed to high temperature processing (Simon et al. 1994). Equilibrium thermodynamic calculations show that $Al_2O_3$ condenses out of a solar-like gas at temperatures between ~1400 – 1600 K, for reasonable photospheric pressures, and, typically, pure stoichiometric spinel condenses at ~200 K below $Al_2O_3$ (Lodders & Fegley 1995; Yoneda & Grossman 1995). If, in fact, the grains formed in a gradually cooling envelope close to the interface temperatures of $Al_2O_3$ and spinel, it may be possible for "extra" Al to be incorporated into spinel, producing the high observed Al/Mg values. Similarly, as suggested by Choi et al. (1998), various stages of partial back-reactions of gas phase Mg atoms with earlier condensed $Al_2O_3$ remaining in the stellar outflow may result in spinel grains with a spread of non-stoichiometric Al/Mg ratios. However, some caveats about assuming that the measured SIMS results for Al/Mg values in small grains are correct should be pointed out here. Quantification of elemental ratios in SIMS analysis, in general, is often problematic, as the elemental composition of the analyzed samples can affect the relative ion yields with respect to one another ("matrix effect") and elemental sensitivity factors must be obtained from measurements on standards of known concentrations. For the spinel grains measured here and in Zinner et al. (2005), however, nearby spinel grains on the sample mount were used for normalization, and as such, there should be no systematic difference. That being said, some previous NanoSIMS studies (e.g., Amari et al. 2002 and unpublished data from our laboratory) have hinted at elemental ratio variations as a function of grain size, in particular for Si/C ratios in small (~0.45 μm – about the same size as the spinel grains in this study) presolar SiC grains. Although we did make our standard measurements on grains of the same mineralogy as the presolar ones, in order to maximize the signal and not have the grains sputter away too quickly, we measured small clumps of grains, or aggregates, and the results of these analyses were used for the calculation of Al/Mg ratios in the presolar grains. If there were a grain size effect, it would certainly affect the situation here as the presolar grains are exceedingly small and the clumps were at least an order of magnitude larger. In addition, as the grain size is roughly the same size as the $O^-$ primary beam rastered over each grain, we cannot absolutely rule out the possibility of some contamination on the sample mount; this may also account for the reason we see slightly elevated Al/Mg ratios even compared to the results of Zinner et al. (2005), although variations in instrumental conditions from session to session can also often affect absolute elemental ratios as well. Though the grains were mounted on gold foil of nominal 99.9999% purity, we have observed in other sample mounts that the gold substrate often contains microscopic specks of $Al_2O_3$ (usually < 100 nm in diameter artifacts from the manufacturer) either on the substrate surface or within it and any contribution from this contamination would certainly increase the measured Al/Mg ratio. It should also be noted that the clumps of grains used for normalization could have contained grains of non-spinel composition inherited from the residue itself, thereby compromising the determination of the Al/Mg sensitivity factor. To possibly resolve this issue, in principle, one could compare the Al/Mg ratios in presolar and solar spinel grains of the same size, although this has yet to be done. It has been pointed out recently (Lin, Gyngard, & Zinner 2010) that the differing ionization efficiencies of Mg and Al (in particular after the implantation of Cs) can affect the Al/Mg ratio if the measurement acquisition time is not long enough to allow the ion signals of the two elements to completely equilibrate. This would be a particular problem when comparing very small grains (for which measurement times are short in order to preserve the grain for future analyses) to calibration spinel grains that have ample material to allow the measurement profiles to sufficiently saturate. In addition, the aggregate grains were not implanted with as much Cs as those that were presolar, as they were not identified and measured by the automatic system. Although overall count rates are generally lower for the individual grains, there do not appear to be significant enough differences in the time required to reach implantation equilibrium of the Mg isotopes and [27]Al between the presolar grains and the nearby spinel grains used for normalization that could account for the elevated Al/Mg ratios in the presolar spinel grains. The [24]Mg, [25]Mg, [26]Mg, and [27]Al ion signals do not show appreciable differences in their profiles as a function of time during the measurements (either on the presolar grains or the aggregates), unlike the situation for small SiC grains – see Figure 3 of (Lin et al. 2010). The Auger spectra taken of both the presolar grains discovered in the automated search and of nearby normal grains (Figure 10) indicate qualitatively similar compositions. Unfortunately, however, difficulty in quantification does not allow us to definitively confirm or refute the Al/Mg ratios as determined by SIMS studies. EDX analysis in the TEM of a focused ion beam lift-out section from several presolar spinel grains have yielded both stoichiometric and non-stoichiometric Al/Mg ratios (Zega et al. 2009, 2010), consistent with the SIMS measurements within uncertainties, though the statistics are limited. Although we cannot absolutely state here that presolar spinel grains are stoichiometric, elevated Al/Mg ratios reported from SIMS studies may be in question, and more investigations are needed to determine the true elemental Al/Mg ratio in larger numbers of presolar spinel grains. Accurate of Al/Mg ratios in presolar spinel could provide information on whether the majority of these grains condensed under the same astrophysical conditions or at a range of temperatures and pressures.

## 5. CONCLUSIONS

In collaboration with Cameca, we developed and implemented a high-mass- and spatial-resolution, automated particle measurement system for the NanoSIMS. As a first test, we successfully applied this technique to a search for presolar spinel grains from the Murray meteorite and discovered 41 O-anomalous grains. The automated system has proven to be robust for efficiently finding and measuring presolar grains and the tight integration of the system into the software interface of the instrument allows for measurement flexibility and software stability. Future improvements on customization, such as the ability to terminate a measurement when a specified statistical precision is achieved or to only perform an individual grain measurement when a specific isotopic anomaly is present in the ion images, are planned in order to broaden the applicability of the technique and increase its efficiency.

For 29 of the grains (26 spinel and 3 $Al_2O_3$), the Mg-Al isotopic systematics were also determined. Most of the data are consistent with previous results on presolar spinel grains, showing slight to moderate enrichments in [25]Mg and often large [26]Mg enhancements, likely due to the decay of [26]Al. At least six grains also have large, strikingly similar [25]Mg depletions and [26]Mg excesses, but different O isotopic compositions, indicative of a combination of GCE of the Mg isotopes and nucleosynthesis in the grains' parent stars. We have identified one spinel grain with large excesses in [17]O, [25]Mg, and [26]Mg that is the best O-rich candidate for being a nova grain discovered to date, and argue that O-rich nova grains are more abundant in primitive meteorites than carbonaceous nova grains. Also, an extreme Group 4 grain and a grain with a large enrichment in [16]O (as well as clear evidence for radiogenic [44]Ca from the decay of [44]Ti), both likely SN condensates, have been found. At least two Group 2 grains have large Mg isotopic anomalies (particularly $\delta^{25}Mg > 150‰$) that are difficult to explain by nucleosynthesis in low-mass AGB stars and may have obtained their Mg isotopic compositions by mass transfer in binary star systems. Future correlated isotopic, elemental, and structural studies of presolar spinel are needed to further identify and characterize the astrophysical origins of these grains and, undoubtedly, automated measurement techniques will be critical to discovering these and other types of rare O-rich grains.

## ACKNOWLEDGEMENTS

We thank Tim Smolar for his tireless efforts in keeping the NanoSIMS functioning, Francois Hillion for facilitating the collaboration with Cameca, and Tom Bernatowicz for helpful discussions. We are grateful to Tang Ming for producing the Murray residue, Roy Lewis for providing the samples, Jordi Jose for assistance with the nova models, and David Fanning for software assistance. This work has been supported by NASA grants NNG05GJ66G (FJS), NNX08AG71G (EZ), and NNX07AJ71G (LRN).

# REFERENCES

Amari, S., Anders, E., Virag, A., & Zinner, E. 1990, Nature, 345, 238

Amari, S., Gao, X., Nittler, L. R., Zinner, E., José, J., Hernanz, M., & Lewis, R. S. 2001, ApJ, 551, 1065

Amari, S., Jennings, C., Nguyen, A., Stadermann, F. J., Zinner, E., & Lewis, R. S. 2002, Lunar Planet. Sci., XXXIII, ed. S. J. Mackwell (League City, TX: LPI), Abstract 1205

Amari, S., Lewis, R. S., & Anders, E. 1994, Geochim. Cosmochim. Acta, 58, 459

Andreä, J., Drechsel, H., & Starrfield, S. 1994, A&A, 291, 869

Bernatowicz, T., Fraundorf, G., Tang, M., Anders, E., Wopenka, B., Zinner, E., & Fraundorf, P. 1987, Nature, 330, 728

Boothroyd, A. I., & Sackmann, I.-J. 1999, ApJ, 510, 232

Boothroyd, A. I., Sackmann, I.-J., & Wasserburg, G. J. 1994, ApJ, 430, L77

Boothroyd, A. I., Sackmann, I.-J., & Wasserburg, G. J. 1995, ApJ, 442, L21

Busso, M., Gallino, R., & Wasserburg, G. J. 1999, ARA&A, 37, 239

Busso, M., Wasserburg, G. J., Nollett, K. M., & Calandra, A. 2007, ApJ, 671, 802

Choi, B.-G., Huss, G. R., Wasserburg, G. J., & Gallino, R. 1998, Science, 282, 1284

Choi, B.-G., Wasserburg, G. J., & Huss, G. R. 1999, ApJ, 522, L133

Clayton, D. D., & Nittler, L. R. 2004, ARA&A, 42, 39

Clayton, R. N. 1993, Ann. Rev. Earth Planet. Sci., 21, 115

Davidson, J., Busemann, H., Alexander, C.M.O'D., Nittler, L. R., Schrader, D. L., Orthous-Daunay, F. R., Quirico, E., Franchi, I. A., & Grady, M. M. 2009, Lunar Planet Sci. XL, ed. S. J. Mackwell (League City, TX: LPI), Abstract 1853

Eggleton, P., Dearborn, D., & Lattanzio, J. 2006, Science, 314, 1580

Floss, C., & Stadermann, F. 2009, Geochim. Cosmochim. Acta, 73, 2415

Gao, X., & Nittler, L. R. 1997, Lunar Planet. Sci., XXVIII, ed. S. J. Mackwell (League City, TX: LPI), Abstract 393

Gehrz, R. D., Truran, J. W., Williams, R. E., & Starrfield, S. 1998, PASP, 110, 3

Gröner, E., & Hoppe, P. 2006, Applied Surface Sci., 252, 7148

Gyngard, F., Morgand, A., Nittler, L. R., Stadermann, F. J., & Zinner, E. 2009, Lunar Planet. Sci., XL, ed. S. J. Mackwell (League City, TX: LPI), Abstract 1386

Gyngard, F., Nittler, L. R., Stadermann, F. J., & Zinner, E. 2010, Lunar Planet. Sci., XLI, ed. S. J. Mackwell (League City, TX: LPI), Abstract 1152

Gyngard, F., & Zinner, E. 2009, Meteorit. Planet. Sci., 44, A82

Heck, P. R., Hoppe, P., Gröner, E., Marhas, K. K., Baur, H., & Wieler, R. 2006, Lunar Planet. Sci., XXXVII, ed. S. J. Mackwell (League City, TX: LPI), Abstract 1355

Heck, P. R., Marhas, K. K., Hoppe, P., Gallino, R., Baur, H., & Wieler, R. 2007, ApJ, 656, 1208

Hernanz, M., José, J., Coc, A., Gómez-Gomar, J., & Isern, J. 1999, ApJ, 526, L97

Hoppe, P., Kocher, T. A., Strebel, R., Eberhardt, P., Amari, S., & Lewis, R. S. 1996, Lunar Planet. Sci., XXVII, ed. S. J. Mackwell (League City, TX: LPI), Abstract 561

Hoppe, P., & Ott, U. 1997, in Astrophysical Implications of the Laboratory Study of Presolar Materials, ed. T. J. Bernatowicz & E. Zinner (New York: AIP), 27

Hughes, J. P., Rakowski, C. E., Burrows, D. N., & Slane, P. O. 2000, ApJ, 528, L109

Hynes, K. M., & Gyngard, F. 2009, Lunar Planet. Sci., XL, ed. S. J. Mackwell (League City, TX: LPI), Abstract 1198

Iliadis, C., Angulo, C., Descouvemont, P., Lugaro, M., & Mohr, P. 2008, Phys. Rev. C, 77, 045802

Iliadis, C., Champagne, A., José, J., Starrfield, S., & Tupper, P. 2002, ApJS, 142, 105

José, J., Coc, A., & Hernanz, M. 2001, ApJ, 560, 897

José, J., & Hernanz, M. 1998, ApJ, 494, 680

José, J., & Hernanz, M. 2007, Meteorit. Planet. Sci., 42, 1135

José, J., Hernanz, M., Amari, S., Lodders, K., & Zinner, E. 2004, ApJ, 612, 414

José, J., Hernanz, M., & Iliadis, C. 2006, Nuc. Phys. A, 777, 550

Kahane, C., Cernicharo, J., Gomez-Gonzalez, J., & Guélin, M. 1992, A&A, 256, 235

Karakas, A. I., & Lattanzio, J. C. 2003, PASA, 20, 279

Kovetz, A., & Prialnik, D. 1997, ApJ, 477, 356

Lattanzio, J. C., & Boothroyd, A. I. 1997, in Astrophysical Implications of the Laboratory Study
    of Presolar Materials, ed. T. J. Bernatowicz & E. Zinner (New York: AIP), 85

Lewis, R. S., Tang, M., Wacker, J. F., Anders, E., & Steel, E. 1987, Nature, 326, 160

Lin, Y., Gyngard, F., & Zinner, E. 2010, ApJ, 709, 1157

Lodders, K. 2003, Apj, 591, 1220

Lodders, K., & Fegley, B., Jr. 1995, Meteoritics, 30, 661

Lugaro, M., Karakas, A. I., Nittler, L. R., Alexander, C. M. O'D., Hoppe, P., Iliadis, C., &
    Lattanzio, J. C. 2007, A&A, 461, 657

Marks, P. B., Sarna, M. J., & Prialnik, D. 1997, MNRAS, 290, 283

Messenger, S., Keller, L. P., & Lauretta, D. S. 2005, Science, 309, 737

Meyer, B. S., Weaver, T. A., & Woosley, S. E. 1995, Meteoritics, 30, 325

Meyer, B. S., & Zinner, E. 2006, in Meteorites and the Early Solar System II, ed. D. S. Lauretta
    & H. Y. McSween, Jr. (Tucson, AZ: Univ. Tucson Press), 69

Nguyen, A. N., & Zinner, E. 2004, Science, 303, 1496

Nguyen, A., Zinner, E., & Lewis, R. S. 2003, PASA, 20, 382

Nguyen, A. N., Stadermann, F. J., Zinner, E., Stroud, R. M., Alexander, C. M. O'D., & Nittler, L. R. 2007, ApJ, 656, 1223

Nittler, L. R. 1996, Quantitative isotopic ratio ion imaging and its application to studies of preserved stardust in meteorites (St. Louis: Washington University)

---. 1997, in Astrophysical Implications of the Laboratory Study of Presolar Materials, ed. T. J. Bernatowicz & E. Zinner (New York: AIP), 59

Nittler, L. R. 2009, PASA, 26, 271

Nittler, L. R., & Alexander, C. M. O'D. 2003, Geochim. Cosmochim. Acta, 67, 4961

Nittler, L. R., Alexander, C. M. O'D., Gallino, R., Hoppe, P., Nguyen, A. N., Stadermann, F. J., & Zinner, E. K. 2008, ApJ, 682, 1450

Nittler, L. R., Alexander, C. M. O'D., Gao, X., Walker, R. M., & Zinner, E. 1994, Nature, 370, 443

Nittler, L. R., Alexander, C. M. O'D., Gao, X., Walker, R. M., & Zinner, E. 1997, ApJ, 483, 475

Nittler, L. R., Alexander, C. M. O'D., & Nguyen, A. N. 2006, Meteorit. Planet. Sci., 41, A134

Nittler, L. R., Alexander, C. M. O'D., Wang, J., & Gao, X. 1998, Nature, 393, 222

Nittler, L. R., & Hoppe, P. 2005, ApJ, 631, L89

Nollett, K. M., Busso, M., & Wasserburg, G. J. 2003, ApJ, 582, 1036

Nozawa, T., Kozasa, T., Habe, A., Dwek, E., Umeda, H., Tominaga, N., Maeda, K., & Nomoto, K. 2007, ApJ, 666, 955

Pagel, B. E. J. 2009, Nucleosynthesis and Chemical Evolution of Galaxies (2nd ed.; Cambridge: Cambridge Univ. Press)

Rauscher, T., Heger, A., Hoffman, R. D., & Woosley, S. E. 2002, ApJ, 576, 323

Simon, S. B., Yoneda, S., Grossman, L., & Davis, A. M. 1994, Geochim. Cosmochim. Acta, 58, 1937

Stadermann, F. J., Floss, C., Bose, M., & Lea, A. S. 2009, Meteorit. Planet. Sci., 44, 1033

Starrfield, S., Iliadis, C., & Hix, W. R. 2008, in Classical Novae, ed. M. F. Bode & A. Evans (Cambridge: Cambridge Univ. Press), 77

Starrfield, S., Truran, J. W., Sparks, W. M., & Kutter, G. S. 1972, ApJ, 176, 169

Starrfield, S., Truran, J. W., Wiescher, M. C., & Sparks, W. M. 1998, MNRAS, 296, 502

Straniero, O., Chieffi, A., Limongi, M., Busso, M., Gallino, R., & Arlandini, C. 1997, ApJ, 478, 332

Stroud, R. M., Nittler, L. R., & Alexander C. M. O'D. 2004, Science, 305, 1445.

Tang, M., & Anders, E. 1988, Geochim. Cosmochim. Acta, 52, 1235

Wasserburg, G. J., Boothroyd, A. I., & Sackmann, I.-J. 1995, ApJ, 447, L37

Woosley, S. E., & Heger, A. 2007, Phys. Rep., 442, 269

Yaron, O., Prialnik, D., Shara, M. M., & Kovetz, A. 2005, ApJ, 623, 398

Yoneda, S., & Grossman, L. 1995, Geochim. Cosmochim. Acta, 59, 3413

Yoshida, T., & Hashimoto, M. 2004, ApJ, 606, 592

Zega, T. J., Alexander, C. M. O'D., Nittler, L. R., & Stroud, R. M. 2009, Lunar Planet. Sci., XL, ed. S. J. Mackwell (League City, TX: LPI), Abstract 1342

---. 2010, Lunar Planet. Sci., XLI, 2055, ed. S. J. Mackwell (League City, TX: LPI), Abstract 2055

Zinner, E. 2007, in Meteorites, Comets and Planets: Treatise on Geochemistry Vol. 1, Update 1, ed. A. M. Davis, H. D. Holland & K. K. Turekian  (Oxford: Elsevier), 1

Zinner, E., Amari, S., Guinness, R., Nguyen, A., Stadermann, F., Walker, R. M., & Lewis, R. S. 2003, Geochim. Cosmochim. Acta, 67, 5083

Zinner, E., & Gyngard, F. 2009, Lunar Planet. Sci., XL, ed. S. J. Mackwell (League City, TX: LPI), Abstract 1046

Zinner, E., Nittler, L. R., Gallino, R., Karakas, A. I., Lugaro, M., Straniero, O., & Lattanzio, J. C. 2006, ApJ, 650, 350

Zinner, E., Nittler, L. R., Hoppe, P., Gallino, R., Straniero, O., & Alexander, C. M. O'D. 2005, Geochim. Cosmochim. Acta, 69, 4149

# FIGURE CAPTIONS

Figure 1:   SEM and O isotopic ratio images of grain 12-13-3 found in this study. a) High-resolution SEM image acquired with the Auger Nanoprobe. b) Low-resolution NanoSIMS secondary electron image of the grain; the lower isotopically normal grain was subsequently sputtered away. c) and d) O isotopic ratio images of $^{17}O/^{16}O$ and $^{18}O/^{16}O$, expressed as delta values (deviations from solar isotopic ratios in parts per thousand, ‰), revealing the top, presolar grain to be a Group 4 grain.

Figure 2:  Plot of the O isotopic compositions of the anomalous grains found in this study and previously identified presolar oxide and silicate grains (see Hynes & Gyngard 2009 and references therein).  The labeled grains are discussed specifically in the text.  In this and all other figures, dashed lines indicate solar-system isotopic ratios and error bars are 1σ.

Figure 3:  Mg isotopic compositions of the presolar grains not sputtered away during O isotopic measurement, expressed as δ-values, or deviations from solar isotopic ratios in parts per thousand (‰).

Figure 4:  Schematic diagram of the stellar processes responsible for the O isotopic composition of most Group 1 – 3 presolar oxide and silicate grains.  Grains C4-8 and T54 plot in the region only accessible by nova nucleosynthesis.  Also shown are ellipses representing the four groups identified for presolar O-rich grains.  Approximate shifts in composition from cool bottom processing (CBP), galactic chemical evolution (GCE), and hot bottom burning (HBB) are given as double black, solid gray, and dotted gray lines, respectively.

Figure 5:  Plot of the O isotopic compositions predicted by various nova models, shown with the data from grains C4-8 and T54.  Errors on the grain data are smaller than their symbol size.  Multiple points for some models are due to variation in the amount of mixing

between the roughly solar accreted material and the WD core. Models explicitly mentioned in the text are labeled in gray.

Figure 6: Plot of the Mg isotopic compositions theoretically predicted for the ejecta of CO and ONe novae (José et al. 2004), compared with the measured Mg isotopic ratios of grain C4-8. For the models, the contribution of [26]Al was also included, multiplied by a factor of 24 to account for the preferential condensation of Al into spinel. Models explicitly mentioned in the text are labeled in gray.

Figure 7: Oxygen 3-isotope plot showing the composition of grain 7-5-7 found in this study, as well as that of $Al_2O_3$ grain T84 (Nittler et al. 1998) and olivine grain B10A (Messenger et al. 2005). Also shown is a mixing line calculated by varying the relative amounts of material from the He/C and O-rich inner zones of the 15 $M_\odot$ SN model of Rauscher et al. (2002).

Figure 8: Theoretical predictions of $^{25}Mg/^{24}Mg$, $^{26}Mg/^{24}Mg$, and $^{26}Al/^{27}Al$ ratios (normalized to solar values) and $^{44}Ti$ abundance from the 15$M_\odot$ SN model of Rauscher et al. (2002).

Figure 9: Plot of calculated isotopic ratios for mixtures of various SN zones for 15$M_\odot$ models from Rauscher et al. (2002) and Woosley and Heger (2007) relative to those of grain 7-5-7.

Figure 10: Two differentiated Auger electron spectra typical of the spinel grains of this study, showing the qualitative agreement between presolar and solar spinel grains. The C and Cs spectral lines are from surface contamination and implanted Cs from the NanoSIMS primary beam, respectively. (a) Presolar spinel grain 6-13-5. (b) A nearby solar-composition grain.

Table 1: Oxygen and Mg-Al isotopic and elemental ratios of the presolar oxides from this work.

| Sample | Group[a] | Mineralogy | $^{17}O/^{16}O$ $(10^{-4})$ | $^{18}O/^{16}O$ $(10^{-3})$ | $\delta^{25}Mg/^{24}Mg$ (‰) | $\delta^{26}Mg/^{24}Mg$ (‰) | Al/Mg[b] |
|---|---|---|---|---|---|---|---|
| 1-2-8 | 1 | $Al_2O_3$ | $10.45 \pm 0.38$ | $1.51 \pm 0.05$ | $-48 \pm 22$ | $2 \pm 22$ | $35.56 \pm 5.36$ |
| 3-3-10 | 1 | Spinel | $13.90 \pm 0.26$ | $1.55 \pm 0.03$ | $-4 \pm 6$ | $-6 \pm 5$ | $2.54 \pm 0.38$ |
| 6-13-5 | 1 | Spinel | $8.93 \pm 0.37$ | $1.73 \pm 0.05$ | $91 \pm 9$ | $211 \pm 9$ | $3.31 \pm 0.50$ |
| 6-14-8 | 1 | Spinel | $7.98 \pm 0.17$ | $2.20 \pm 0.03$ | $-235 \pm 20$ | $216 \pm 25$ | $5.17 \pm 0.78$ |
| 6-24-10 | 1 | Spinel | $10.17 \pm 0.21$ | $1.31 \pm 0.03$ | $-220 \pm 5$ | $309 \pm 5$ | $2.56 \pm 0.39$ |
| 7-1-4 | 1 | Spinel | $7.10 \pm 0.38$ | $2.04 \pm 0.07$ | $-190 \pm 9$ | $255 \pm 10$ | $3.28 \pm 0.49$ |
| 7-5-7[d] | U | Spinel | $0.40 \pm 0.10$ | $0.21 \pm 0.02$ | $-276 \pm 10$ | $13 \pm 12$ | $3.09 \pm 0.46$ |
| 8-24-13 | 1 | Spinel | $14.85 \pm 0.70$ | $1.96 \pm 0.08$ | $43 \pm 9$ | $91 \pm 8$ | $3.06 \pm 0.46$ |
| 8-9-3 | 1 | $Al_2O_3$ | $51.42 \pm 1.05$ | $1.89 \pm 0.07$ | $-66 \pm 21$ | $-25 \pm 20$ | $40.04 \pm 6.04$ |
| 9-10-17 | 2 | Spinel | $11.54 \pm 0.42$ | $0.64 \pm 0.04$ | $-191 \pm 8$ | $376 \pm 10$ | $3.38 \pm 0.51$ |
| 9-18-8 | 1 | Spinel | $4.96 \pm 0.21$ | $1.81 \pm 0.04$ | $-8 \pm 6$ | $-3 \pm 5$ | $2.84 \pm 0.43$ |
| 9-20-15 | 1 | Spinel | $5.37 \pm 0.24$ | $1.64 \pm 0.04$ | $7 \pm 7$ | $10 \pm 6$ | $2.53 \pm 0.38$ |
| 9-21-11 | 1 | Spinel | $13.90 \pm 0.42$ | $2.01 \pm 0.05$ | $-12 \pm 5$ | $-3 \pm 4$ | $2.51 \pm 0.38$ |
| 9-27-18 | 1 | Spinel | $5.31 \pm 0.20$ | $1.96 \pm 0.04$ | $0 \pm 12$ | $16 \pm 11$ | $3.18 \pm 0.48$ |
| 9-8-20 | 2 | Spinel | $12.87 \pm 0.56$ | $0.56 \pm 0.04$ | $-186 \pm 8$ | $341 \pm 9$ | $3.17 \pm 0.48$ |
| 10-11-4 | 3 | Spinel | $3.54 \pm 0.09$ | $1.41 \pm 0.02$ | $-9 \pm 5$ | $-8 \pm 5$ | $3.00 \pm 0.45$ |
| 11-1-1 | 1 | Spinel | $9.50 \pm 0.27$ | $1.56 \pm 0.04$ | $-42 \pm 7$ | $-29 \pm 6$ | $2.63 \pm 0.40$ |
| 12-12-4 | 1 | Spinel | $6.39 \pm 0.36$ | $1.23 \pm 0.05$ | $-23 \pm 14$ | $9 \pm 14$ | $3.68 \pm 0.55$ |
| 12-13-3 | 4 | Spinel | $11.16 \pm 0.52$ | $6.62 \pm 0.13$ | $71 \pm 13$ | $735 \pm 16$ | $3.69 \pm 0.56$ |
| 13-28-12 | 1 | Spinel | $4.82 \pm 0.32$ | $1.92 \pm 0.07$ | $24 \pm 10$ | $15 \pm 10$ | $2.86 \pm 0.43$ |
| 13-30-9 | 2 | Spinel | $11.89 \pm 0.32$ | $0.07 \pm 0.02$ | $150 \pm 7$ | $602 \pm 8$ | $3.11 \pm 0.47$ |
| 13-34-3 | 2 | Spinel | $10.09 \pm 0.32$ | $0.49 \pm 0.03$ | $-183 \pm 7$ | $354 \pm 8$ | $2.87 \pm 0.43$ |
| 13-6-3 | 1 | Spinel | $4.64 \pm 0.37$ | $2.18 \pm 0.08$ | $-46 \pm 24$ | $-20 \pm 23$ | $2.26 \pm 0.34$ |
| 14-1-14 | 2 | $Al_2O_3$ | $9.88 \pm 0.38$ | $0.37 \pm 0.03$ | $-1 \pm 33$ | $727 \pm 43$ | $111.12 \pm 16.77$ |
| 14-12-7 | 2 | Spinel | $7.65 \pm 0.24$ | $0.23 \pm 0.02$ | $1034 \pm 8$ | $1040 \pm 8$ | $2.39 \pm 0.36$ |
| 14-13-6 | 1 | Spinel | $4.45 \pm 0.31$ | $1.40 \pm 0.06$ | $-7 \pm 8$ | $12 \pm 8$ | $3.48 \pm 0.52$ |
| 14-14-5 | 1 | Spinel | $6.51 \pm 0.23$ | $1.34 \pm 0.04$ | $1 \pm 8$ | $-6 \pm 8$ | $2.86 \pm 0.43$ |
| 14-19-2 | 1 | Spinel | $13.67 \pm 0.27$ | $1.96 \pm 0.04$ | $1 \pm 6$ | $-3 \pm 5$ | $3.12 \pm 0.47$ |
| C4-8 | U | Spinel | $440.36 \pm 1.22$ | $1.10 \pm 0.02$ | $949 \pm 8$ | $929 \pm 8$ | $2.89 \pm 0.43$ |
| 3-6-8 | 1 | Spinel | $5.05 \pm 0.30$ | $1.65 \pm 0.06$ | | | |
| 6-16-1 | 1 | Spinel | $7.20 \pm 0.16$ | $1.83 \pm 0.03$ | | | |
| 6-2-1 | 1 | Spinel | 7.17 ± 0.14 | 1.81 ± 0.03 | | | |
| 6-21-9 | 1 | Spinel | 6.61 ± 0.15 | 1.33 ± 0.03 | | | |
| 8-17-2 | 1 | Spinel | 5.94 ± 0.29 | 1.12 ± 0.05 | | | |
| 8-9-11 | 1 | Spinel | 4.54 ± 0.10 | 1.87 ± 0.03 | | | |
| 9-23-6 | 1 | Spinel | 9.76 ± 1.17 | 2.30 ± 0.18 | | | |
| 10-7-16 | 1 | Spinel | 9.15 ± 0.21 | 0.98 ± 0.03 | | | |
| 12-20-10 | U | Spinel | 88.01 ± 2.97 | 1.18 ± 0.11 | | | |
| 13-3-12 | 1 | Spinel | 5.83 ± 0.55 | 2.07 ± 0.10 | | | |
| 14-6-3 | 1 | Spinel | 5.25 ± 0.21 | 1.84 ± 0.04 | | | |
| 9-28-4 | 1 | Spinel | 8.69 ± 0.26 | 1.86 ± 0.04 | | | |
| Solar | | | 3.83 | 2.01 | ≡ 0 | ≡ 0 | 0.082[c] |

Note – All errors are 1σ.  Grain C4-8 was not discovered by the automated measurement system.

[a] U signifies "Unusual".

[b] Determined NanoSIMS analysis; errors are from both counting statistics and reproducibility on the standards.

[c] Lodders, 2003.

[d] $^{44}Ti/^{48}Ti = (3.6 \pm 0.8) \times 10^{-3}$.

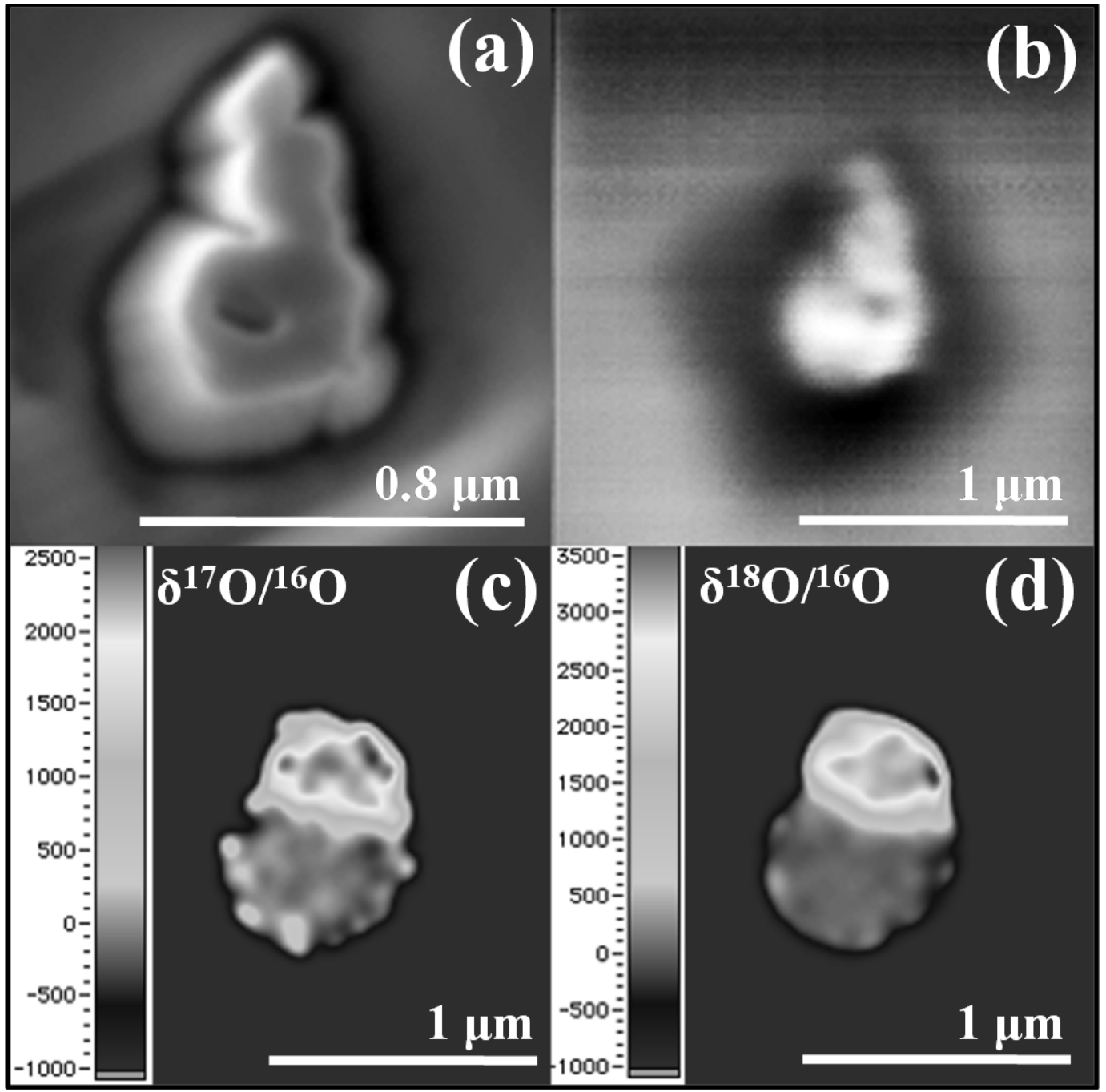

**Figure 1 (f1.eps)**

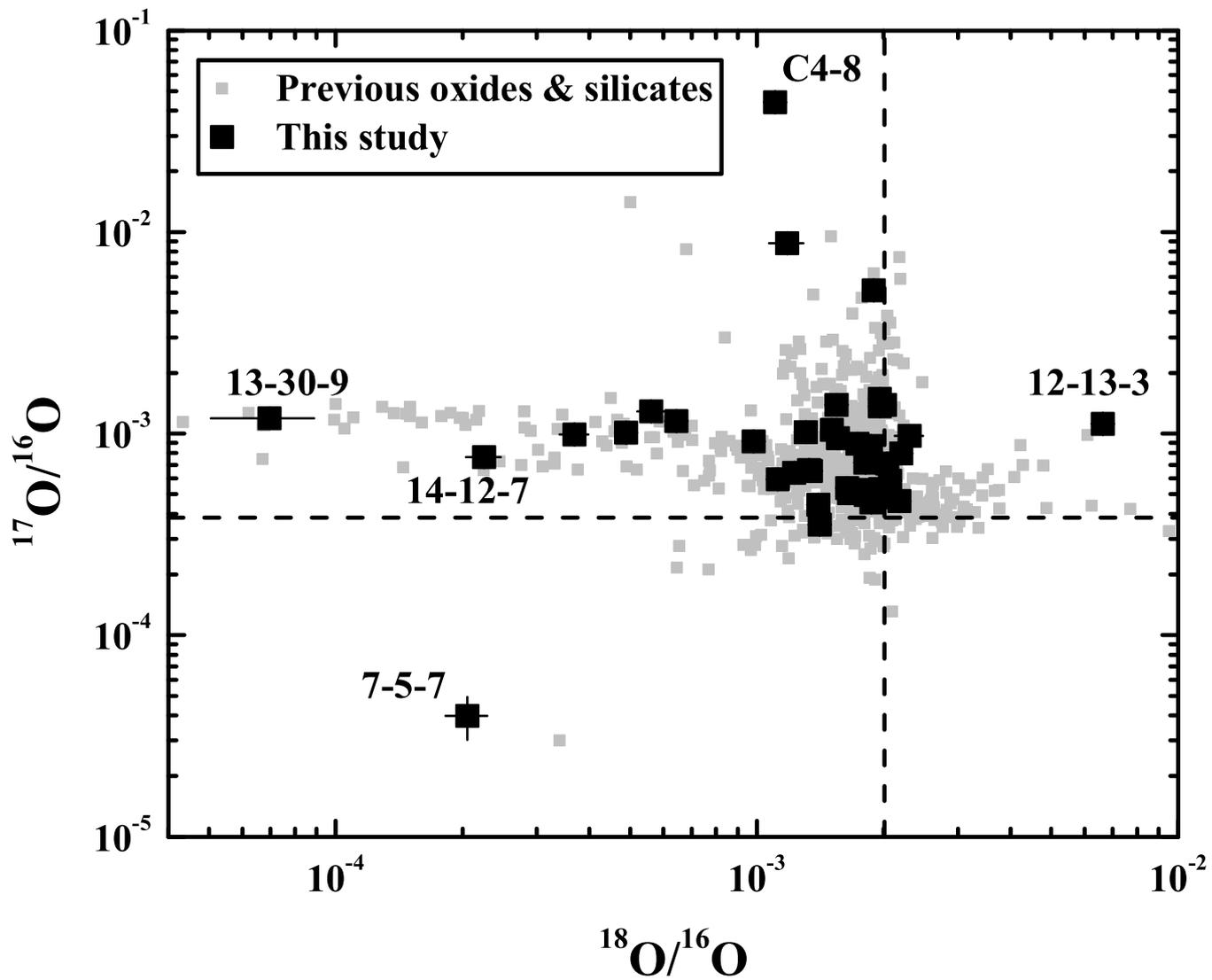

Figure 2 (f2.eps)

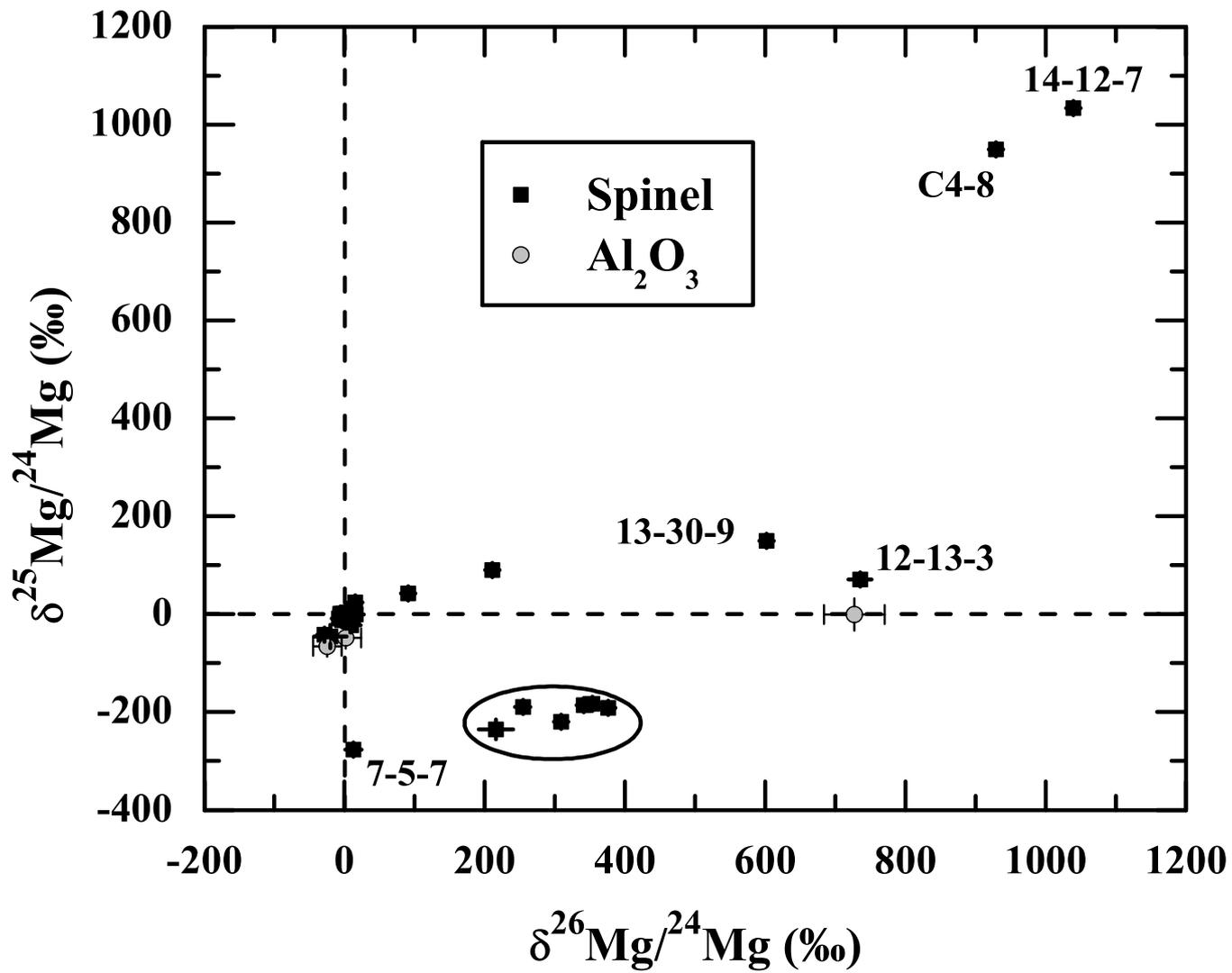

Figure 3 (f3.eps)

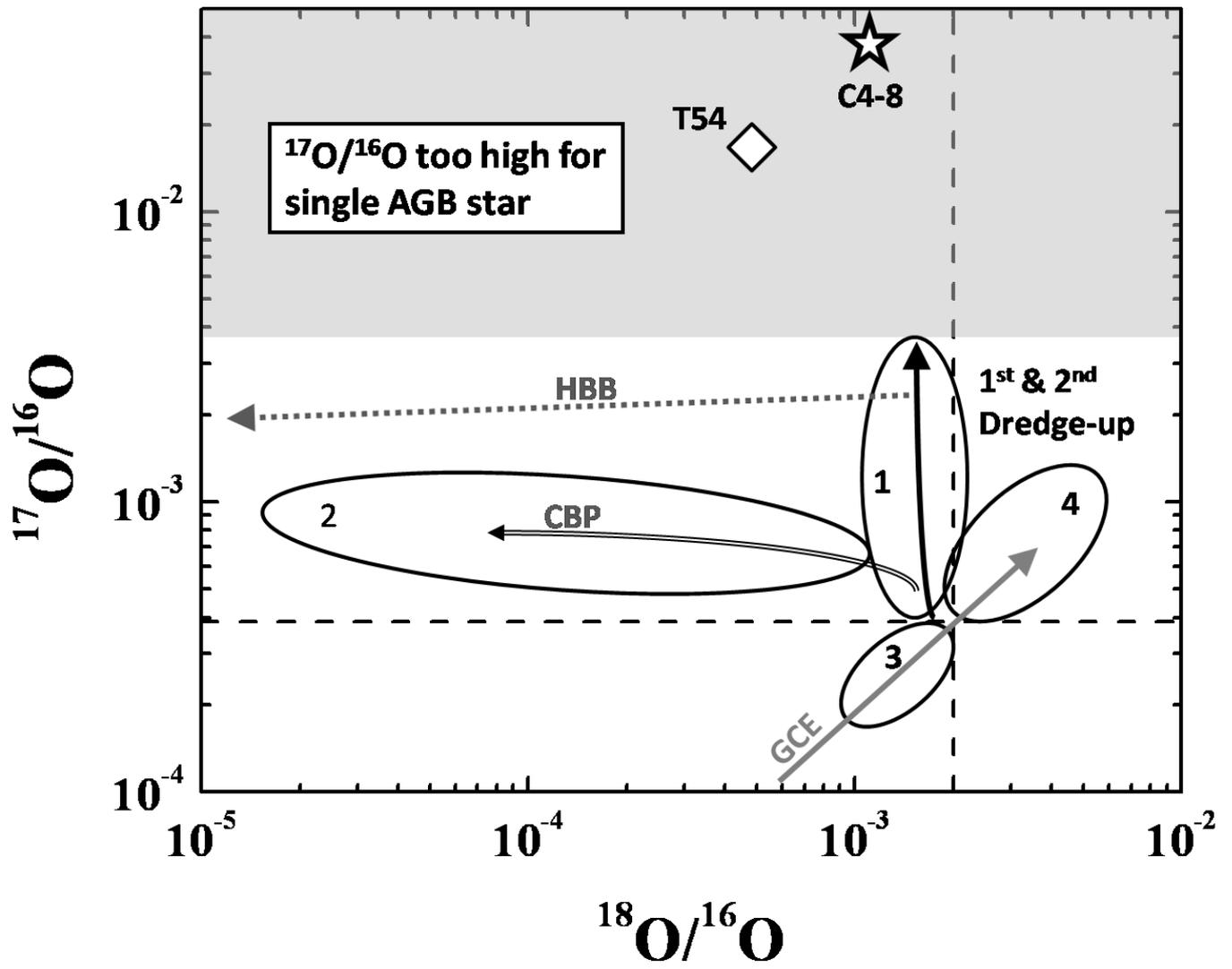

Figure 4 (f4.eps)

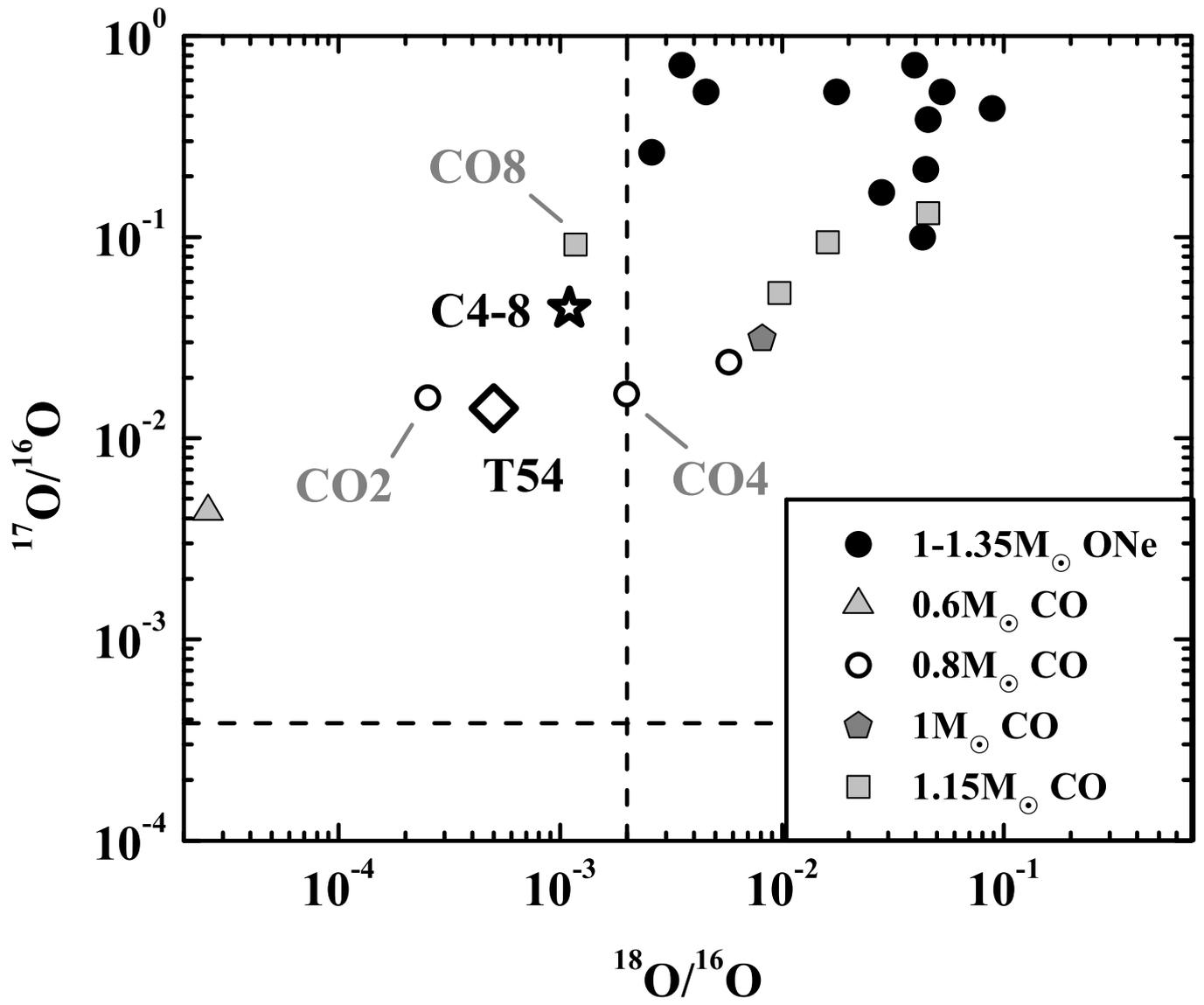

**Figure 5 (f5.eps)**

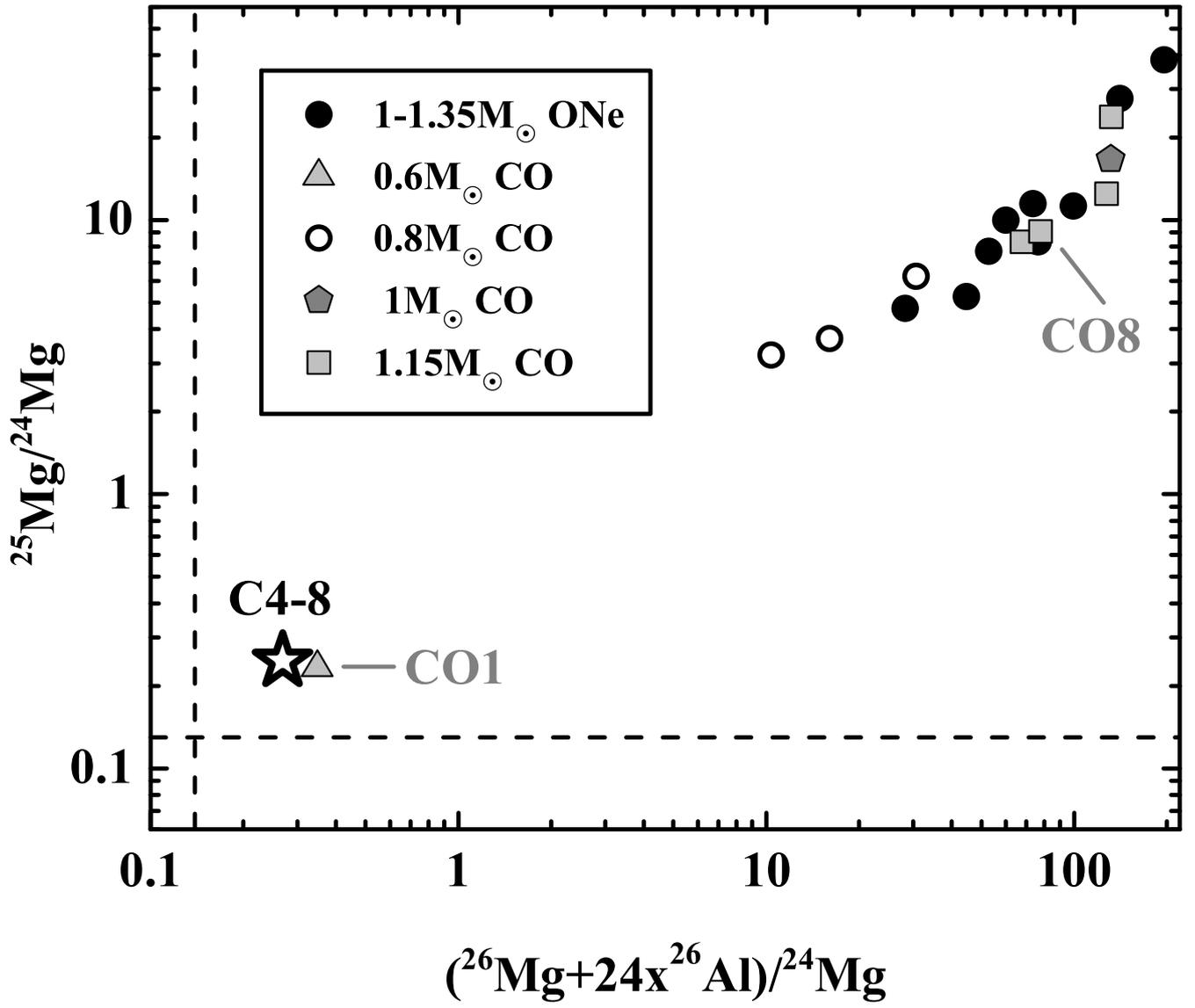

Figure 6 (f6.eps)

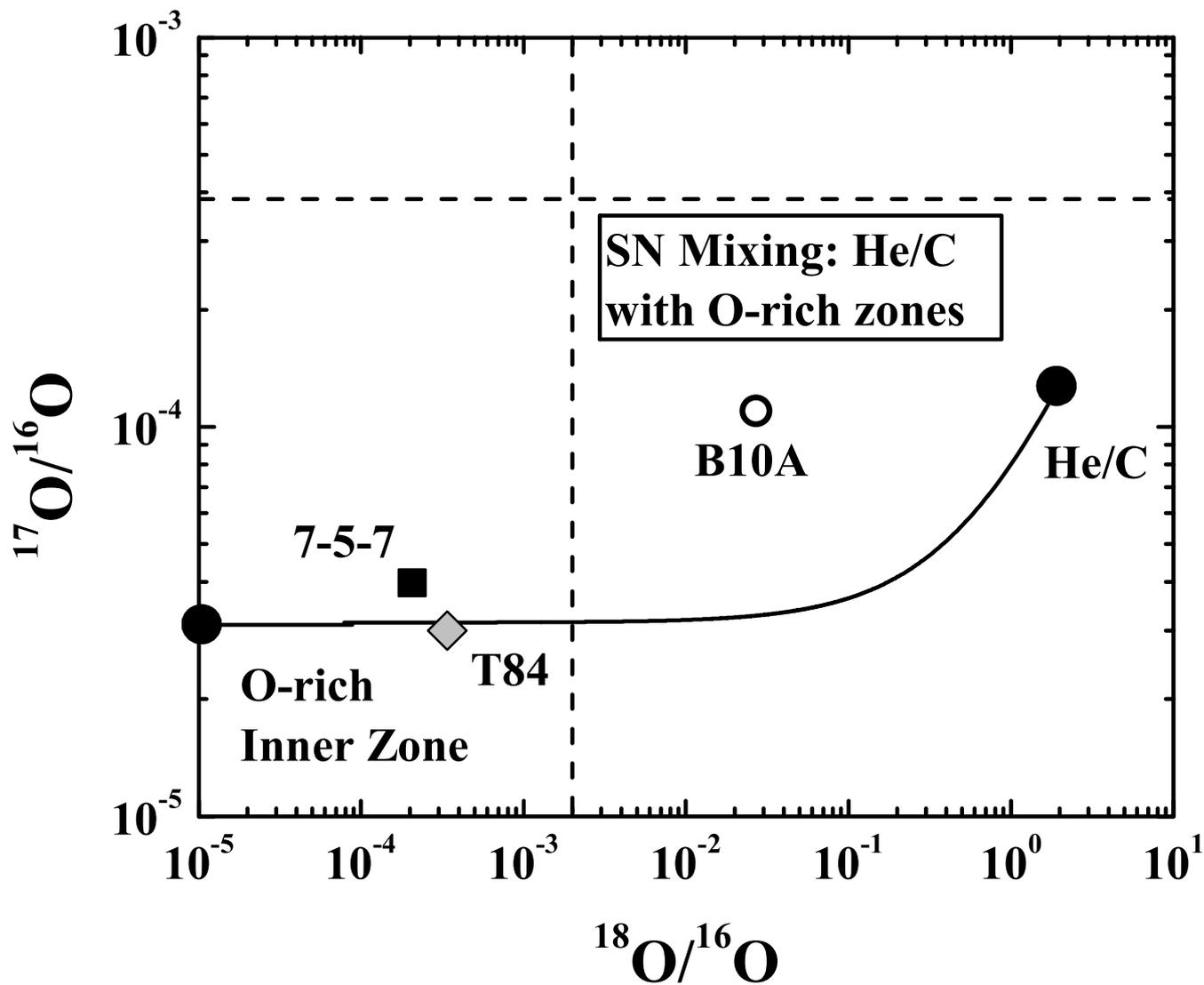

Figure 7 (f7.eps)

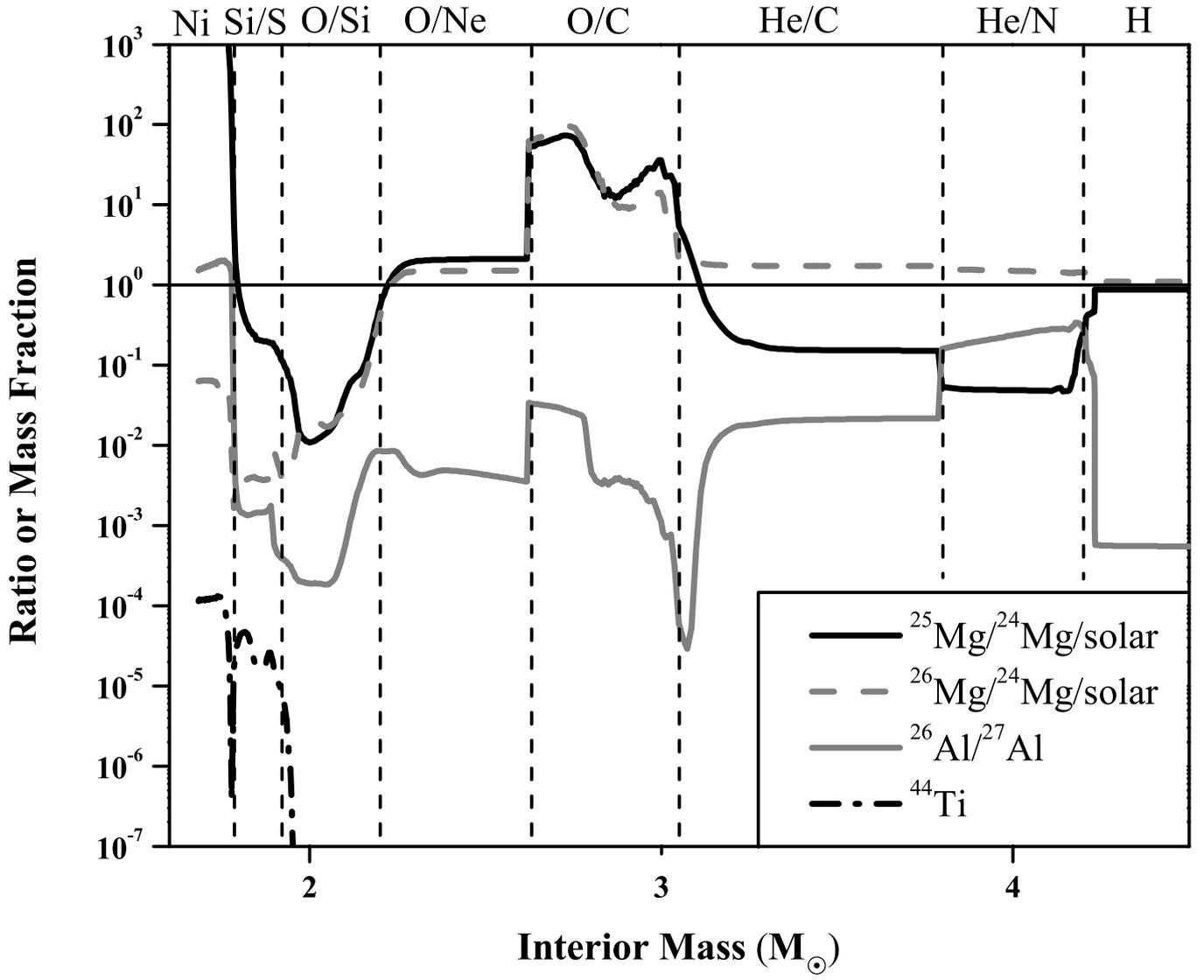

**Figure 8 (f8.eps)**

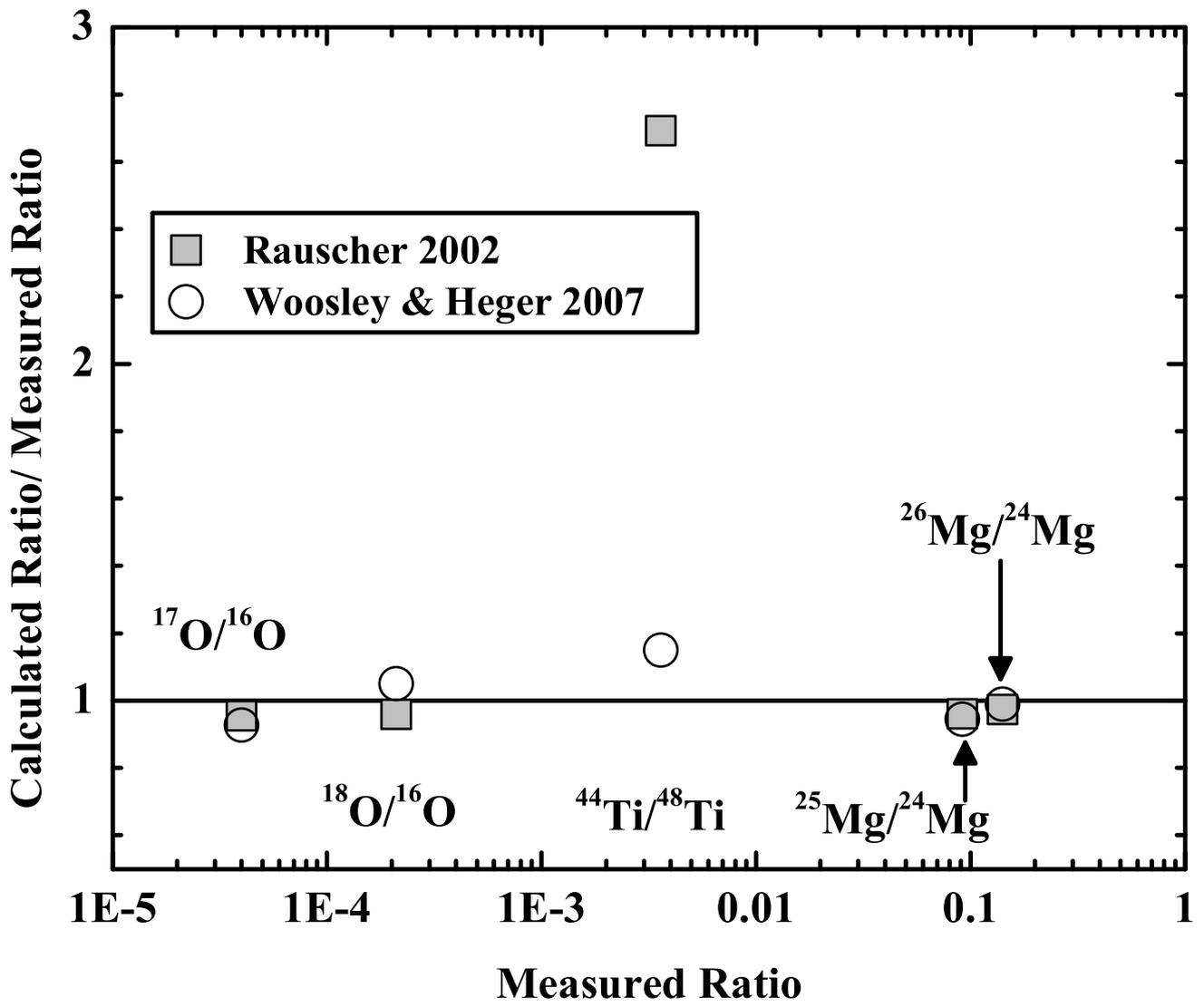

Figure 9 (f9.eps)

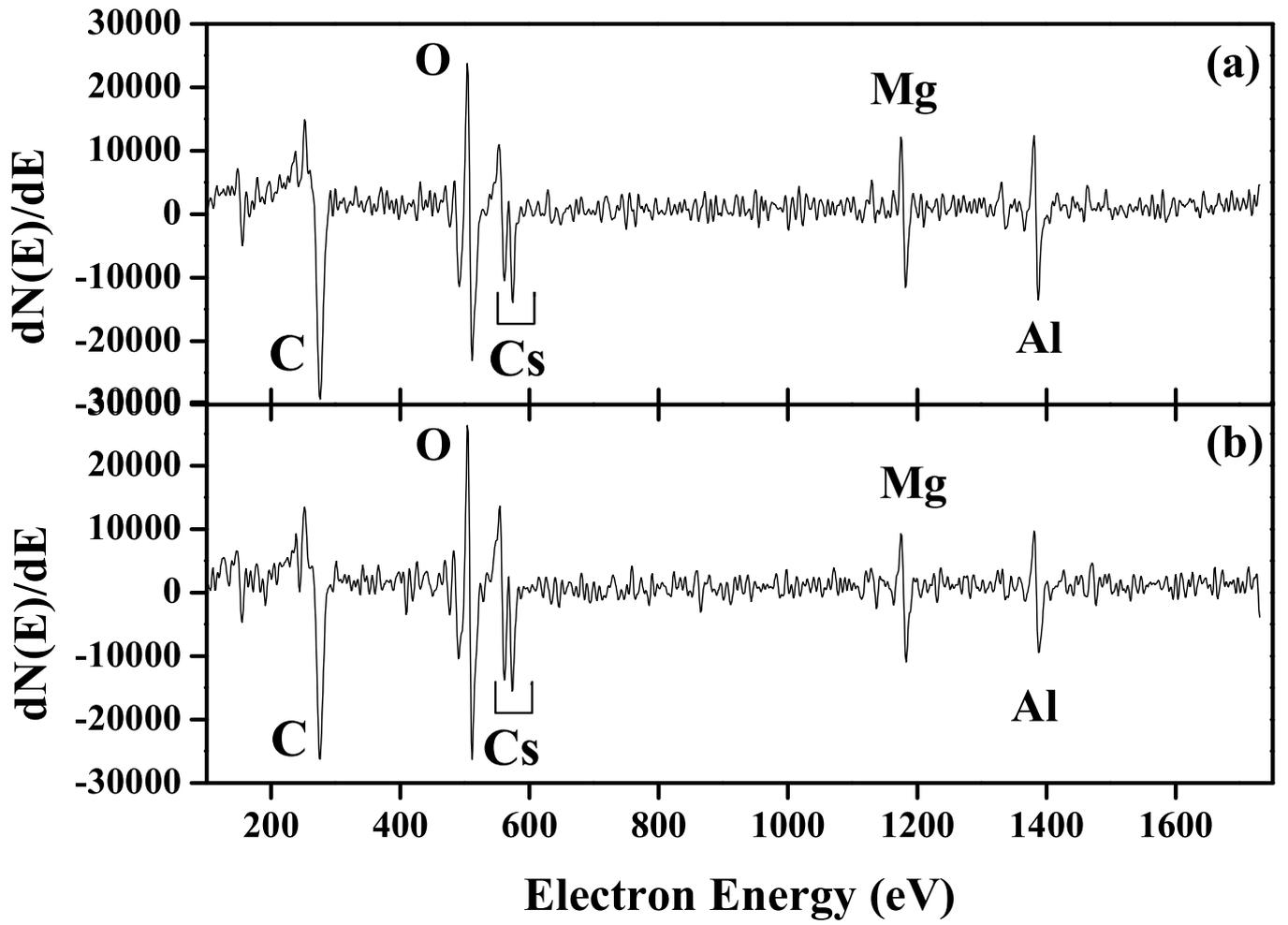

**Figure 10 (f10.eps)**